\documentclass[modern, mathlines]{aastex7}

\usepackage{mdwlist}
\usepackage{hyperref}
\usepackage{breqn}

\begin{document}

\title{Data Release 1 of the Dark Energy Spectroscopic Instrument}


\collaboration{308}{DESI Collaboration}

\author[0009-0000-7133-142X]{M.~Abdul Karim}
\email{marie-lynn.abdulkarim@cea.fr}
\affiliation{IRFU, CEA, Universit\'{e} Paris-Saclay, F-91191 Gif-sur-Yvette, France}

\author{A.~G.~Adame}
\email{julian.adamek@uzh.ch}
\affiliation{Instituto de F\'{\i}sica Te\'{o}rica (IFT) UAM/CSIC, Universidad Aut\'{o}noma de Madrid, Cantoblanco, E-28049, Madrid, Spain}

\author[0000-0001-5200-3973]{D.~Aguado}
\email{david.aguado@iac.es}
\affiliation{Instituto de Astrof\'{\i}sica de Canarias, C/ V\'{\i}a L\'{a}ctea, s/n, E-38205 La Laguna, Tenerife, Spain}

\author{J.~Aguilar}
\email{jaguilar@lbl.gov}
\affiliation{Lawrence Berkeley National Laboratory, 1 Cyclotron Road, Berkeley, CA 94720, USA}

\author[0000-0001-6098-7247]{S.~Ahlen}
\email{ahlen@bu.edu}
\affiliation{Physics Dept., Boston University, 590 Commonwealth Avenue, Boston, MA 02215, USA}

\author[0000-0002-3757-6359]{S.~Alam}
\email{shadab.alam@tifr.res.in}
\affiliation{Tata Institute of Fundamental Research, Homi Bhabha Road, Mumbai 400005, India}

\author{G.~Aldering}
\email{galdering@lbl.gov}
\affiliation{Lawrence Berkeley National Laboratory, 1 Cyclotron Road, Berkeley, CA 94720, USA}

\author[0000-0002-5896-6313]{D.~M.~Alexander}
\email{d.m.alexander@durham.ac.uk}
\affiliation{Centre for Extragalactic Astronomy, Department of Physics, Durham University, South Road, Durham, DH1 3LE, UK}
\affiliation{Institute for Computational Cosmology, Department of Physics, Durham University, South Road, Durham DH1 3LE, UK}

\author{R.~Alfarsy}
\email{rahma.alfarsy@port.ac.uk}
\affiliation{Institute of Cosmology and Gravitation, University of Portsmouth, Dennis Sciama Building, Portsmouth, PO1 3FX, UK}

\author{L.~Allen}
\email{lori.allen@noirlab.edu}
\affiliation{NSF NOIRLab, 950 N. Cherry Ave., Tucson, AZ 85719, USA}

% ...

\shorttitle{DESI DR1}
\shortauthors{DESI Collaboration et~al.}
\submitjournal{\aj}

\begin{abstract}
In 2021 May the Dark Energy Spectroscopic Instrument (DESI) collaboration began a 5-year spectroscopic redshift survey to produce a detailed map of the evolving three-dimensional structure of the universe between $z=0$ and $z\approx4$. DESI's principle scientific objectives are to place precise constraints on the equation of state of dark energy, the gravitationally driven growth of large-scale structure, and the sum of the neutrino masses, and to explore the observational signatures of primordial inflation. We present DESI Data Release 1 (DR1), which consists of all data acquired during the first 13 months of the DESI main survey, as well as a uniform reprocessing of the DESI Survey Validation data which was previously made public in the DESI Early Data Release. The DR1 main survey includes high-confidence redshifts for 18.7M objects, of which 13.1M are spectroscopically classified as galaxies, 1.6M as quasars, and 4M as stars, making DR1 the largest sample of extragalactic redshifts ever assembled. We summarize the DR1 observations, the spectroscopic data-reduction pipeline and data products, large-scale structure catalogs, value-added catalogs, and describe how to access and interact with the data. In addition to fulfilling its core cosmological objectives with unprecedented precision, we expect DR1 to enable a wide range of transformational astrophysical studies and discoveries.
\end{abstract}

\setcounter{footnote}{0} 

\section{Introduction}\label{sec:intro}

\subsection{The First Year of DESI Data}\label{sec:firstyear}

Elucidating the nature of dark energy and the physical mechanisms responsible for the accelerating expansion of the universe is one of the most important unsolved problems in physics and, arguably, all of science \citep{albrecht06a, weinberg13a}. To tackle this question, in 2021 May the Dark Energy Spectroscopic Instrument (DESI) collaboration began a 5-year survey to produce the most detailed three-dimensional map of the universe ever assembled \citep{Snowmass2013.Levi, DESI2016a.Science, DESI2016b.Instr}. By measuring the baryon acoustic oscillation (BAO) feature and redshift space distortions at multiple cosmological epochs, DESI aims to place unprecedented constraints on the equation of state of dark energy, the gravitationally driven growth of large-scale structure, and the sum of the neutrino masses, as well as to explore the observational signatures of primordial inflation \citep{DESI2023a.KP1.SV}. Concurrently, DESI is carrying out an ambitious survey of stars in the halo of the Milky Way Galaxy in order to constrain the geometry, properties, and accretion history of its stellar halo, disk, and dark-matter halo \citep{MWS.TS.Cooper.2023}. Notably, DESI has the distinction of being the first Stage-IV dark-energy experiment to begin science operations \citep{albrecht06a}.

\begin{figure}[t]
\centering 
\includegraphics[width=1\columnwidth]{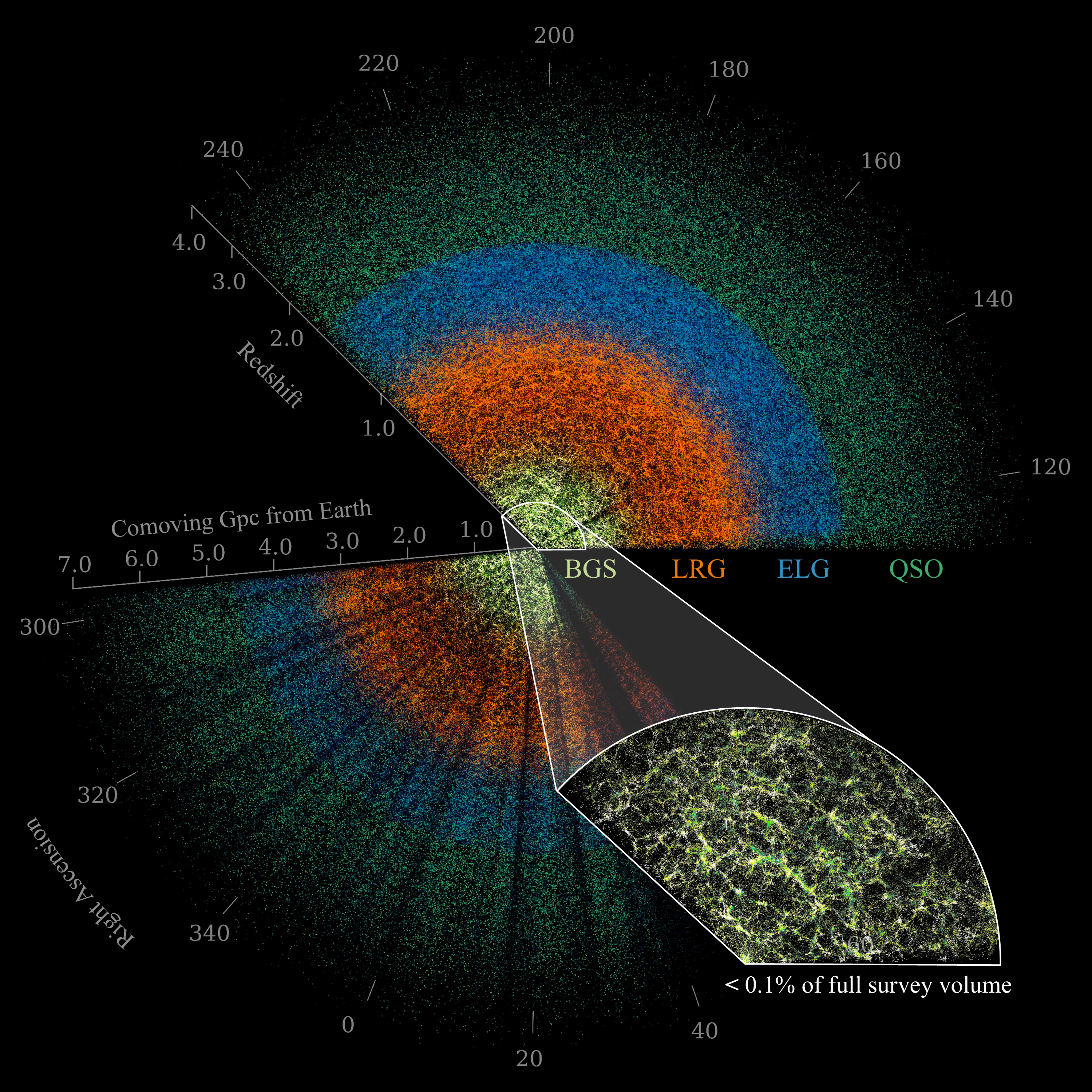}
\caption{A slice of the universe mapped by DR1 drawn from a small wedge of the DESI footprint between $\pm5\arcdeg$ in declination out to $z\approx4$. We render the four major extragalactic samples---bright galaxy survey (BGS) galaxies, luminous red galaxies (LRG), emission-line galaxies (ELG), and QSOs---using yellow, orange, blue, and green points, respectively. Within each target class, the shade of the color maps to declination (lighter colors correspond to higher declination). The inset shows a subset of the BGS survey extending out to redshift $z=0.2$, highlighting the large-scale structure traced by galaxies in the densest survey region. For reference, this small wedge of the BGS survey represents less than 0.1\% of the comoving cosmological volume in DR1. Also note the black streaks of apparent missing points (most visible at right ascensions between 40$\arcdeg$--300$\arcdeg$), which are due to incomplete survey coverage in DR1 which will be populated in future data releases (see \S\ref{sec:sample}). \label{fig:pie}}
\end{figure}

DESI is a highly multiplexed instrument mounted at the prime focus of the Mayall 4-meter telescope at Kitt Peak National Observatory (KPNO) in Arizona, USA. Its 5000 robotic fibers and $3\fdg2$ diameter field-of-view enable it to rapidly acquire optical spectrophotometry of tens of thousands of targets per night \citep{DESI2022.KP1.Instr, SurveyOps.Schlafly.2023}. By the end of its 5-year survey in 2026 May, current projections are that DESI will have measured precise redshifts for approximately 50M unique galaxies and quasars and spectroscopic properties of 25M Milky Way stars.

DESI identifies its primary spectroscopic targets using 14,000~deg$^{2}$ of broadband optical and mid-infrared photometry from Data Release~9 (DR9) of the DESI Legacy Imaging Surveys\footnote{\url{https://www.legacysurvey.org/dr9}} (\citealt{BASS.Zou.2017, LS.Overview.Dey.2019, SGA.Moustakas.2023, TS.Pipeline.Myers.2023}; hereafter, the Legacy Surveys), and, in some cases, using Gaia stellar photometry \citep{gaia-collaboration16a}. Specifically, DESI targets five broad classes of objects scaffolded in redshift: Milky Way survey (MWS) and backup program stars \citep{MWS.TS.Cooper.2023, dey25a}; bright galaxy survey (BGS) galaxies \citep[$0<z<0.6$;][]{BGSPrelim.RuizMacias.2020, BGS.TS.Hahn.2023}; luminous red galaxies \citep[LRGs, $0.4<z<1.1$;][]{LGRPrelim.Zhou.2020, LRG.TS.Zhou.2023}; emission-line galaxies \citep[ELGs, $0.6<z<1.6$;][]{ELGPrelim.Raichoor.2020, ELG.TS.Raichoor.2023}; and QSOs \citep[$0.9<z<4$;][]{QSOPrelim.Yeche.2020, QSO.TS.Chaussidon.2023}. For its cosmological analyses, DESI further differentiates QSO targets into ``tracer'' QSOs at lower redshift, $z<2.1$, from Ly$\alpha$ forest QSOs at $z>2.1$, because above $z=2.1$ DESI uses the Ly$\alpha$ forest as an independent probe of the matter-density field \citep{DESI2016a.Science}.

\begin{figure}[t]
\centering 
\includegraphics[width=1\columnwidth]{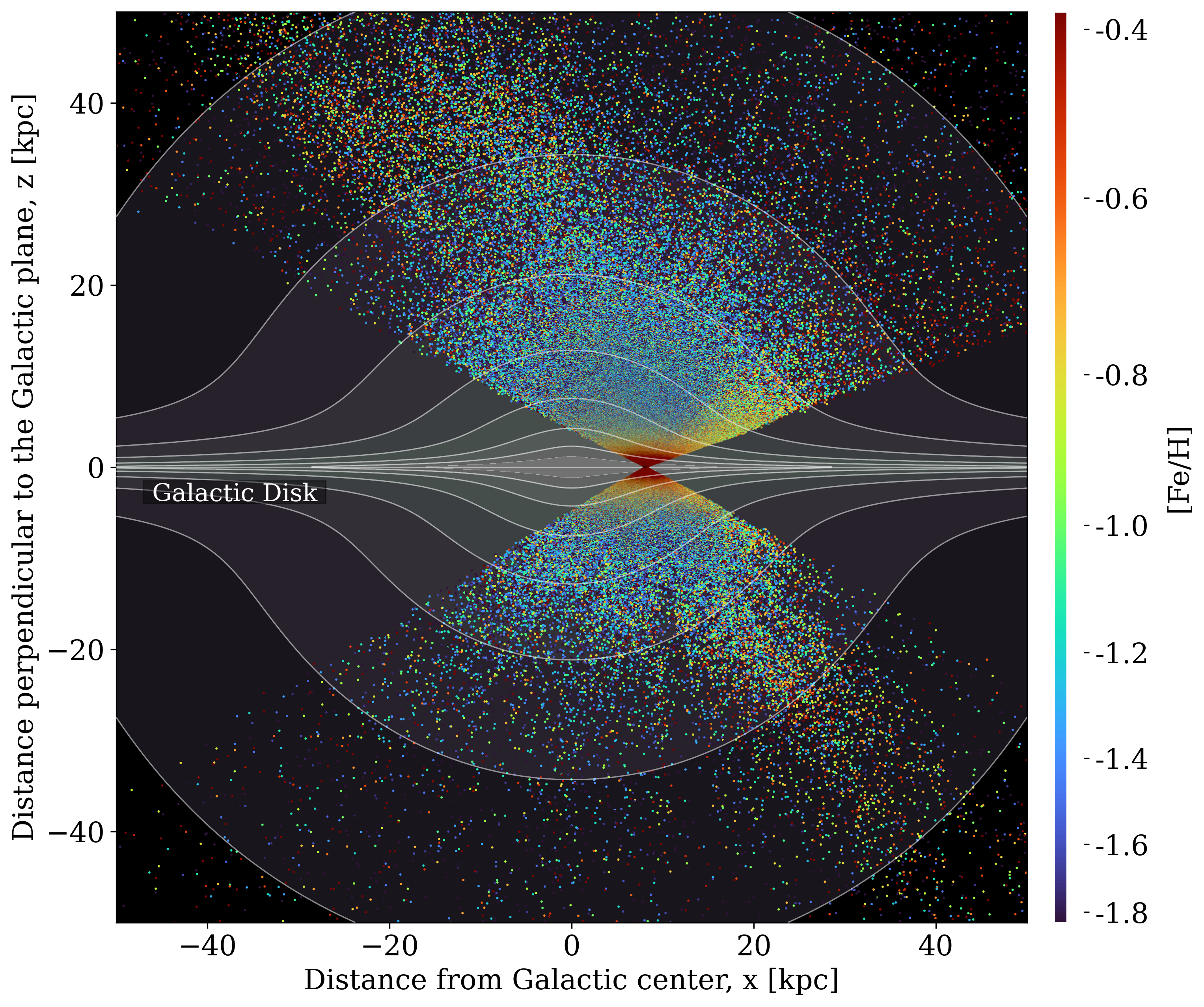}
\caption{Distribution of Milky Way stars in DESI DR1 observed as part of the Milky Way survey. The colors of individual points represent the spectroscopically inferred iron abundance, [Fe/H], as measured using the RVSPecFit pipeline \citep[see Appendix~\ref{app:vacs} and][]{koposov24a}. The distances to individual stars have been derived using a neural network with stellar parameters as inputs. The thin curves represent the disk stellar mass density contours from the \citet{price-whelan17a} \texttt{MilkyWayPotential2022} Galactic model. This visualization illustrates the phenomenal size and scale of the DESI stellar survey, the clear negative iron abundance gradient (from the inner disk to the outer stellar halo), as well as the presence of the Sagittarius stream \citep{majewski03a}, visible as more orange-tinted points at $(x,z)\approx (-10,+40)$~kpc and $(x,z)\approx(20,-20$)~kpc above and below the Galactic disk, respectively. The map also shows high-metallicity stars extending  above the disk at $x=20$~kpc caused by the Monoceros ``stream" structure \citep{juric08a, newberg02a}. \label{fig:mws-pie}}
\end{figure}

After first light and a brief commissioning period, in 2020 December DESI began a five-month survey validation (SV) program which was designed to test the performance of the instrument and all its subsystems, and to validate its target-selection algorithms, data-reduction pipeline, and scientific deliverables \citep{DESI2023a.KP1.SV}. SV consisted of three successive phases: Target Selection Validation (SV1); Operations Development (SV2); and the One-Percent Survey (SV3), the last of which covered roughly 1\% of the final 14,000~deg$^{2}$ DESI footprint but to higher target completeness \citep{DESI2023a.KP1.SV}. These observations, particularly the One-Percent Survey data, resulted in a flurry of scientific activity\footnote{\url{https://data.desi.lbl.gov/doc/papers/edr}} as well as DESI's first measurement of the BAO peak in the BGS, LRG, ELG, and QSO 2-point correlation functions \citep{BAO.EDR.Moon.2023} and in the auto-correlation function of the Ly$\alpha$ forest \citep{gordon23a}. Subsequently, on 2021 May 14, DESI launched its 5-year main survey.\footnote{In detail, a small number of SV observations continued until 2021 June 10, but those data were acquired independently of the main survey and are included in the SV dataset \citep{DESI2023b.KP1.EDR}.}

All the data obtained during SV were first made public on 2023 June 13 as part of the DESI Early Data Release \citep[EDR;][]{DESI2023a.KP1.SV, DESI2023b.KP1.EDR}.\footnote{\url{https://data.desi.lbl.gov/doc/releases/edr}} Here, we present DESI Data Release 1 (DR1), which includes all the data obtained by DESI during its first 13 months of science operations (2021 May 14 through 2022 June 13) as well as a uniform reprocessing of all the SV data. Overall, DESI has been efficient, running well-ahead of schedule and occasionally obtaining redshifts for more than one million unique targets in a single month \citep{SurveyOps.Schlafly.2023}. In the first year of the main survey, DESI has measured confident redshifts for approximately 18.7M unique targets over more than $9,000$~deg$^{2}$, including 13.1M galaxies, 1.6M quasars, and 4M stars, making DESI the largest extragalactic spectroscopic redshift survey ever conducted. For comparison, all five generations of the Sloan Digital Sky Survey \citep[SDSS I/II, III, IV, and V;][]{york00a, dawson13a, majewski17a, blanton17a, kollmeier17a}, spanning 18 public data releases and approximately 25 years of operations,\footnote{\url{https://www.sdss.org}} have cumulatively resulted in approximately 4M unique extragalactic spectra, making DESI DR1 nearly a factor of four larger than all previous SDSS programs combined.



Figures~\ref{fig:pie} and \ref{fig:mws-pie} illustrate the incredible scope and detail of the data included in DESI DR1. In Figure~\ref{fig:pie} we show a two-dimensional projection of redshift and right ascension for a narrow wedge of the DESI footprint ($\pm5$~degrees in declination), unveiling the large-scale structure traced by the BGS, LRG, ELG, and QSO targets out to $z\approx4$. In Figure~\ref{fig:mws-pie} we show a complementary visualization of the Galactic disk mapped out by DESI, showing the physical distances of millions of MWS stars as a function of iron abundance, [Fe/H]. All the data used to generate these two figures are being released in DR1.

In \S\ref{sec:guide} we outline the organization and contents of this paper and describe several important terms and concepts which will help orient readers interested in using DESI DR1. Complementarily, we recommend that readers interested in the cosmological results from DESI DR1 begin at the DESI portal\footnote{\url{https://data.desi.lbl.gov/doc/papers/\#year-1-cosmology-results}} for a high-level overview of the results before consulting the individual key papers \citep{DESI2024.II.KP3, DESI2024.III.KP4, DESI2024.IV.KP6, DESI2024.V.KP5, DESI2024.VI.KP7A, DESI2024.VII.KP7B}.

\subsection{A High-Level Guide to DESI and Data Release 1}\label{sec:guide}

The sheer size and scope of the DESI survey and data can be overwhelming to both new and expert users. The goal of this section is to introduce some of the key, high-level DESI concepts (and vernacular), with the overarching goal of demystifying some of this complexity. All the terminology and concepts we describe here will also apply to subsequent data releases, thereby lowering the barrier to working with future public DESI datasets.

All DESI observations are made in the context of a \textit{survey} and a \textit{program} \citep{SurveyOps.Schlafly.2023}. As introduced in \S\ref{sec:firstyear}, DESI's primary, 5-year scientific program is its main survey, which was preceded by three independent, successive surveys (SV1, SV2, and SV3), collectively called SV.
In addition, an object can be observed as part of the \textit{special} survey, which includes a miscellaneous assortment of targets and observations which are kept separate from the main survey (see \S\ref{sec:special}).

Within each survey, objects are observed in one of three possible programs according to the acceptable lunar phase and on-sky observing conditions (see \S\ref{sec:main}): \textit{bright}, \textit{dark}, or \textit{backup}.\footnote{A fourth program name, \textit{other}, is occassionally used for ad hoc or bespoke observations which do not easily fit into the three standard program names.} By design, LRG, ELG, and QSO targets are assigned to the dark program and BGS and MWS targets are assigned to the bright program; meanwhile, the backup program consists entirely of stellar (Milky Way) targets, and special (secondary and tertiary targets) can be assigned to any program \citep[see \S\ref{sec:targets};][]{SurveyOps.Schlafly.2023, TS.Pipeline.Myers.2023, DESI2023a.KP1.SV}. In general, objects within each survey/program combination are always treated independently, from target selection (\S\ref{sec:targets}) all the way through to spectral coaddition and redshift estimation (\S\ref{sec:datareduction}); in other words, the same astrophysical object can appear in two (or more) survey/program combinations. This strict separation of targets within each survey and program is a critical part of the large-scale structure pipeline which enables DESI to make precise cosmological measurements \citep{KP3s15-Ross}.

The next important DESI concept is the spectroscopic production, or \texttt{specprod}, which is a fixed or pre-defined set of input spectra and calibration files processed with a well-defined (tagged) software stack. The philosophy underpinning these productions is that the provenance of every file and measurement is well-defined and fully reproducible. In DESI, spectroscopic productions are alphabetically named after mountains or mountain ranges. Notably, a given data release such as DR1 may contain more than one \texttt{specprod} (and also note that not all spectroscopic productions are publicly released since some are for internal testing and validation).

\begin{deluxetable}{lcccc}[t]
\tablecaption{Summary of DESI Data Releases\label{tab:drsummary}}
\tablehead{
\colhead{Spectroscopic} & \colhead{Software} & \colhead{Included} & \colhead{No. of Unique} & \colhead{Release} \\
\colhead{Production\tablenotemark{a}} & \colhead{Stack} & \colhead{Surveys\tablenotemark{b}} & \colhead{Redshifts\tablenotemark{c}} & \colhead{Date}
}
\startdata
\multicolumn{5}{c}{\textbf{Early Data Release (EDR)}\tablenotemark{d}} \\
\hline
\hline
Fuji & \texttt{fuji}\tablenotemark{e} & SV & 1.7M & 2023 June \\
\hline
\multicolumn{5}{c}{\textbf{Data Release 1 (DR1)}\tablenotemark{f}} \\
\hline
\hline
Guadalupe\tablenotemark{g} & \texttt{fuji}\tablenotemark{e} & Main (2 months) & 2.4M & 2025 March \\
Iron & \texttt{iron}\tablenotemark{h} & SV, Main (13 months) & 20.4M & 2025 March \\
\enddata
\tablenotetext{a}{Within the DESI data model, a spectroscopic production is referred to as a \texttt{specprod}. Each \texttt{specprod} corresponds to a uniform processing of the data using a well-defined and reproducible set of input files and software tags.}
\tablenotetext{b}{See \S\ref{sec:operations} for a description of the SV and main surveys.}
\tablenotetext{c}{Approximate total number of unique redshifts summed over all surveys and programs.}
\tablenotetext{d}{The Early Data Release is described in \citet{DESI2023b.KP1.EDR} and the data are publicly available at \url{https://data.desi.lbl.gov/doc/releases/edr}.}
\tablenotemark{e}{\url{https://data.desi.lbl.gov/doc/releases/edr/software-version}}
\tablenotetext{f}{Some papers based on DESI data will interchangeably refer to DR1 as the ``Year 1'' (Y1) DESI sample; in addition, unless explicitly stated otherwise, DR1 (or Y1) implicitly means ``the Iron spectroscopic production in DR1.''}
\tablenotetext{g}{Some early DESI papers refer to the Guadalupe production as ``M2'' or ``DESI-M2'', referencing the two months of main-survey data.}
\tablenotemark{h}{\url{https://data.desi.lbl.gov/doc/releases/dr1/software-version}}
\end{deluxetable}

The primary spectroscopic production for DR1 is Iron, named after the Iron Mountain in Utah, USA; the Iron spectroscopic production includes data from the first 13 months of the DESI main survey, as well as a uniform reprocessing of the SV data which was previously released in the EDR \citep{DESI2023b.KP1.EDR}. DR1 also includes a supplemental or ancillary spectroscopic production called Guadalupe (after the Guadalupe Mountains in Texas, USA), which includes just the first two months of the DESI main survey data (see Appendix~\ref{app:guadalupe}). We include Guadalupe in DR1 because it was used in some early DESI analyses \citep[e.g.,][]{BAO.EDR.Moon.2023, gordon23a, ravoux23a, karacayli24a, herrera-alcantar23a, KP6s4-Bault, ramirez-perez24a}. Guadalupe was processed using the same set of calibration files and software tags as Fuji, the \texttt{specprod} used for the EDR \citep{DESI2023b.KP1.EDR}. We emphasize, however, that all new or planned analyses of DESI data should use Iron instead of Guadalupe, since Iron contains significantly more data than Guadalupe and was processed with better algorithms and calibration files (see \S\ref{sec:datareduction}).

Table~\ref{tab:drsummary} summarizes the connection between the public data releases (EDR and DR1) and the spectroscopic productions included in these releases. Unless otherwise noted, in this paper we use DR1 to mean ``observations, files, and measurements based on the Iron spectroscopic production,'' including all summary statistics and performance metrics. Finally, we point out that some DESI papers, including most of the cosmological analysis papers \citep[e.g.,][]{DESI2024.VI.KP7A, DESI2024.VII.KP7B} use Year 1 (or Y1) to mean the DR1 or Iron spectroscopic production. So for all intents and purposes, DR1, Y1, and Iron can be used interchangeably to refer to the same underlying set of DESI data products.

We conclude the introduction by outlining the remaining structure of the paper. In \S\ref{sec:data}, we provide an overview of DESI as an instrument (\S\ref{sec:instrument}), describe some of the key DESI target-selection concepts (\S\ref{sec:targets}), and summarize the observations contained in DR1 and the status of the DESI main survey (\S\ref{sec:operations}). Next, \S\ref{sec:redux} describes how DESI derives redshifts and spectral classifications from the raw data (\S\ref{sec:datareduction}); tabulates and visualizes the number and distribution of unique, confident redshifts in DR1 (\S\ref{sec:sample}); describes the major DR1 data products and how the data are organized (\S\ref{sec:products}); summarizes the large-scale structure catalogs used in the DESI Y1 cosmological analyses (\S\ref{sec:lss}); and presents the current set of DR1 value-added catalogs (\S\ref{sec:vac-files}). Finally, in \S\ref{sec:access} we describe how the public data can be accessed using two complementary interfaces, and in \S\ref{sec:summary} we summarize the paper.

\section{Data Acquisition}\label{sec:data}

In this section we document how DESI targets and observes objects over its $14,000$~deg$^{2}$ footprint. In \S\ref{sec:instrument} we describe the characteristics of DESI as a highly multiplexed spectroscopic instrument; in \S\ref{sec:targets} we briefly summarize how primary, secondary, and tertiary targets are selected for observations; and in \S\ref{sec:operations} we outline DESI survey operations, the process by which DESI carries out its observing program, and summarize the observations contained within DR1.

\subsection{Dark Energy Spectroscopic Instrument}\label{sec:instrument}

DESI is capable of observing 5000 objects simultaneously using 10 petals of 500 fibers each over an 8~deg$^{2}$ field-of-view focal plane \citep{FocalPlane.Silber.2023, Corrector.Miller.2023}. These fibers send light to 10 corresponding spectrographs with three arms, or cameras, sensitive to a different portion of the 3600--9800~\AA{} observed-frame wavelength range \citep{FiberSystem.Poppett.2024}. The three cameras are  sensitive to blue, red, and near-infrared light, and are denoted B, R, and Z, respectively. In its nominal configuration, DESI generates ten $4096\times4096$-pixel blue images and twenty $4114\times4128$-pixel red and near-infrared images, each containing data for 500 fibers. Table~\ref{tab:instrument_parameters} summarizes some of the other key parameters of the instrument; additional details can be found in \citet{DESI2022.KP1.Instr} and \citet{Spectro.Pipeline.Guy.2023}.

The DESI wavelength coverage and instrumental resolution are designed to resolve the [\ion{O}{2}]~$\lambda\lambda3726,29$ doublet for galaxies at redshifts $0.6<z<1.6$, with the throughput optimized to measure [\ion{O}{2}] down to fluxes of $8\times10^{-17}~\text{erg}~\text{s}^{-1}~\text{cm}^{-2}$ with an effective exposure time of 1000~s. Here, effective exposure time corresponds to an exposure time under nominal dark-time conditions---an object observed at zenith, in a dark sky with full-width at half-maximum (FWHM) seeing of 1\farcs1, and no Galactic extinction (see \S4.14 of \citealt{Spectro.Pipeline.Guy.2023} for additional details).

\begin{deluxetable}{cccc}[t]
\tablecaption{Key DESI (Instrument) Parameters \label{tab:instrument_parameters}}
\tablehead{\multicolumn{2}{c}{Parameter} & \multicolumn{2}{c}{Value}}
\startdata
\multicolumn{2}{l}{Field of View} & \multicolumn{2}{c}{8~deg${^2}$} \\
\multicolumn{2}{l}{Number of spectrographs} & \multicolumn{2}{c}{10} \\
\multicolumn{2}{l}{Fibers per spectrograph} & \multicolumn{2}{c}{500} \\
\multicolumn{2}{l}{Cameras per spectrograph} & \multicolumn{2}{c}{3 (B,R,Z)} \\
\hline
\hline
Camera & Spectral Range & Pixel Size & Resolution \\
Name   & (\AA)          & (\AA)      & ($\lambda/\Delta\lambda$) \\
\hline
  B    & 3600--5800     & 0.8 & 2000--3500 \\
  R    & 5760--7620     & 0.8 & 3400--4800 \\
  Z    & 7520--9824     & 0.8 & 3800--5200 \\
\enddata
\tablecomments{See \citet{DESI2022.KP1.Instr} and \citet{Spectro.Pipeline.Guy.2023} for a detailed description of DESI.}
\end{deluxetable}

\subsection{Target Selection}\label{sec:targets}

\citet{DESI2023a.KP1.SV} and \citet{TS.Pipeline.Myers.2023} provide a summary of the selection criteria DESI uses to identify targets for spectroscopic observations, including estimates of the completeness, contamination rate, and technical implementation (building on a significant body of work documented in \citealt{DESI2016a.Science}, and references therein). In addition, individual supporting papers delve into the target-selection algorithms for each specific class of objects targeted by DESI, which we summarize in Table~\ref{tab:tspapers} \citep[updated from a prior version of this table presented in][]{TS.Pipeline.Myers.2023}. Here, we briefly describe some key target-selection concepts and terminology used elsewhere in the paper, but defer to the individual papers cited in Table~\ref{tab:tspapers} for additional details regarding DESI target selection.

Targets selected as part of the initial DESI design specifications \citep[BGS, ELG, LRG, QSO, and MWS targets;][]{DESI2016a.Science} are typically referred to as \textit{primary} targets. Primary targets are always photometrically selected from the Legacy Surveys or from Gaia (\citealt{gaia-collaboration16a}; see \citealt{TS.Pipeline.Myers.2023} for details). The photometric selection criteria include carefully tuned color-cuts designed to maximize the number density and redshift efficiency of each target class, in addition to the following magnitude limits: $r < 19.5$~mag (BGS\_BRIGHT); $r < 20.175$~mag (BGS\_FAINT); $z_{\text{fiber}} < 21.6$~mag (LRGs); $g_{\text{fiber}} < 24.1$~mag (ELGs); and $r < 23$~mag (QSOs). In Table~\ref{tab:tspapers} we list the references which justify and document the complete set of magnitude and color-selection criteria adopted for each class (see also \citealt{TS.Pipeline.Myers.2023}).

MWS and BGS targets are observed in bright observing conditions while LRG, ELG, and QSO targets are observed in dark conditions (see \S\ref{sec:main} and \citealt{SurveyOps.Schlafly.2023} for a quantitative definition of bright and dark observing conditions). In addition, DESI observes stellar \textit{backup} targets during twilight, or when conditions are too poor for main-survey observations; these backup targets are considered part of the MWS primary program (Dey et~al. 2025, in preparation).

\begin{deluxetable*}{llll}[t]
\tablecaption{Summary of DESI Target-Selection Publications}\label{tab:tspapers}
\tablewidth{0pt}
\tablehead{
\colhead{Target Selection} &
\colhead{Main Survey} &
\colhead{} &
\colhead{} \\
\colhead{Category} &
\colhead{Bit Name\tablenotemark{a}} &
\colhead{Description} &
\colhead{Reference}
}
\startdata
\sidehead{\textit{Bright-time targets}}
 ~~~Bright Galaxy Survey & \texttt{BGS\_ANY} & Any BGS bit is set & \citet{BGS.TS.Hahn.2023} \\
                      & \texttt{BGS\_FAINT\tablenotemark{b}} & Faint BGS target & \\
                      & \texttt{BGS\_BRIGHT} & Bright BGS target & \\
                      & \texttt{BGS\_WISE} & AGN-like BGS target & \citet{juneau25a} \\
                      & \texttt{BGS\_FAINT\_HIP} & Faint BGS target prioritized & \\
                      &        & like a bright BGS target\tablenotemark{c} & \\
~~~Milky Way Survey & \texttt{MWS\_ANY} & Any MWS bit is set & \citet{MWS.TS.Cooper.2023}\\
                  & \texttt{MWS\_BROAD}\tablenotemark{d} & Magnitude-limited bulk sample &  \\
                  & \texttt{MWS\_WD} & White dwarf &  \\
                  & \texttt{MWS\_NEARBY} & Volume-limited $\sim$100\,pc sample &  \\
                  & \texttt{MWS\_BHB} & Blue Horizontal Branch target &  \\
                  & \texttt{MWS\_MAIN\_BLUE} & Magnitude-limited blue sample &  \\
                  & \texttt{MWS\_MAIN\_RED} & Magnitude-limited red sample &  \\
\sidehead{\textit{Dark-time targets}}
 ~~~Luminous Red Galaxies & \texttt{LRG} & LRG target & \citet{LRG.TS.Zhou.2023} \\
 ~~~Emission Line Galaxies\tablenotemark{e} & \texttt{ELG} & ELG target & \citet{ELG.TS.Raichoor.2023} \\
                        & \texttt{ELG\_LOP} & ELG at standard priority & \\
                        & \texttt{ELG\_HIP} & ELG observed at the (higher) & \\
                        &                & priority of an LRG & \\
                        & \texttt{ELG\_VLO} & Low-priority ``filler'' ELG & \\
 ~~~Quasars & \texttt{QSO} & Quasar target & \citet{QSO.TS.Chaussidon.2023} \\
\sidehead{\textit{Backup targets}}
 ~~~~~~~~Part of the & \texttt{BACKUP\_GIANT\_LOP}\tablenotemark{f}    & Candidate Giant Star & \citet{dey25a} \\
  ~~~Milky Way Survey &                                             & observed at lower priority  &  \\
     & \texttt{BACKUP\_GIANT}        & Candidate Giant Star &  \\
                          & \texttt{BACKUP\_BRIGHT}   & Brighter backup target &  \\
                          & \texttt{BACKUP\_FAINT}        & Fainter backup target &  \\
                          & \texttt{BACKUP\_VERY\_FAINT}  & Even fainter backup target &  \\
\enddata
\tablenotetext{a}{Stored as bit-values in the \texttt{DESI\_TARGET}, \texttt{BGS\_TARGET} and \texttt{MWS\_TARGET} columns. Bit-names can be converted to bit-values using the \texttt{desi\_mask}, \texttt{bgs\_mask} and \texttt{mws\_mask} bitmasks \citep[see \S2.4 in][]{TS.Pipeline.Myers.2023}.}
\tablenotetext{b}{BGS bits other than \texttt{BGS\_ANY} are stored in the \texttt{BGS\_TARGET} column and \texttt{bgs\_mask} bitmask.}
\tablenotetext{c}{Some targets with low observational priority are observed at higher priority to help characterize the survey selection function.}
\tablenotetext{d}{MWS bits other than \texttt{MWS\_ANY} are stored in the \texttt{MWS\_TARGET} column and \texttt{mws\_mask} bitmask.}
\tablenotetext{e}{Every ELG is also assigned bits from combinations of ELG\_LOP, ELG\_HIP, or ELG\_VLO. See \S3 of \citet{ELG.TS.Raichoor.2023} for further details.}
\tablenotetext{f}{BACKUP bits are stored in the \texttt{MWS\_TARGET} column and \texttt{mws\_mask} bitmask.}
\end{deluxetable*}

In addition to primary targets, DESI also targets \textit{secondary} and \textit{tertiary} targets, in order to facilitate bespoke science programs with goals beyond the primary DESI experiment \citep[e.g.,][]{palmese21a, darragh-ford23a, dey23a, saulder23a, yang23a, fawcett23a, soumagnac24a, manser24a, bystrom25a, valluri25a}. Secondary targets can be selected using any source of imaging, although the vast majority of these targets have counterparts in the Legacy Surveys. Tertiary targets are similar to secondary targets, but they are observed on dedicated \textit{special} survey tiles instead of being interleaved with regular targets on normal tiles. 

Targets identified for spectroscopic observations are carefully tracked for downstream redshift and large-scale structure catalogs (\citealt{KP3s15-Ross}) using a unique \texttt{TARGETID} derived from its sky position and the imaging data release from which it was selected; consequently, cross-matching across observations or catalogs is performed exclusively by \texttt{TARGETID} rather than by sky position, ensuring unambiguous identification of the same astrophysical object across all DESI observations (see \S3 of \citealt{TS.Pipeline.Myers.2023} for details on the \texttt{TARGETID} construction).

In addition, we record the provenance of each target using names and values stored in dedicated bit-masks (see \S2 of \citealt{TS.Pipeline.Myers.2023} and Table~\ref{tab:tspapers}). In Appendix~\ref{app:primarytargets}, \ref{app:secondarytargets} and \ref{app:tertiarytargets} we summarize the bits used to track primary, secondary and tertiary targets observed as part of the DESI main survey, respectively; the bits used to track SV target classes are documented in the appendices of the EDR paper, \citet{DESI2023b.KP1.EDR}.

\subsection{Survey Operations}\label{sec:operations}

In this section we describe DESI survey operations, the process by which the survey is planned and executed on timescales ranging from a single night to months and years \citep{SurveyOps.Schlafly.2023}. A key concept for all DESI observations is the ``tile,'' which refers to a particular assignment of DESI's fibers to celestial targets. With the exception of some early, unguided commissioning observations, whenever DESI observes, it observes a tile. Each tile is assigned a unique \texttt{TILEID}, although note that multiple exposures of the same \texttt{TILEID} can be executed (each one assigned a unique \texttt{EXPID}). Thus, when discussing the different programs observed by DESI, we frequently refer to the number of tiles and the amount of time spent on those tiles.

In \textsection\ref{sec:main} we describe main-survey operations, which DESI focused on almost exclusively in the time period covered by DR1, 2021 May 14 through 2022 June 13. Observations of SV tiles were included in the EDR and are described in \citet{DESI2023b.KP1.EDR, DESI2023a.KP1.SV}, although we summarize these observations briefly in \textsection\ref{sec:sv} since they are being re-released in DR1. Finally, \textsection\ref{sec:special} briefly introduces the set of observations taken as part of the special survey included in DR1.

\subsubsection{The Main Survey}\label{sec:main}

The DESI main survey consists of 9929 dark tiles, 5676 bright tiles, and 2657 backup tiles, which DESI aims to observe over the course of its 5-year survey \citep{SurveyOps.Schlafly.2023}.\footnote{As the DESI survey is performing better than anticipated, the dark- and bright-time programs have been extended---both in terms of the number of tiles and the exact footprint---since the start of science operations. However, in this paper we adopt the survey definition in place at the start of the survey, as documented in \citet{SurveyOps.Schlafly.2023}, since it is that survey that was used in the DESI year-one cosmology analyses.} For the dark- and bright-time surveys, the locations of the centers of these tiles are chosen to cover the Galactic caps accessible to DESI, roughly the $-23\fdg5 < \mathrm{Dec} < 77\fdg7$ sky where Legacy Surveys imaging is available \citep{LS.Overview.Dey.2019}, supplemented with modest additional cuts to remove tiles at low Galactic latitude (roughly $|b| > 20\arcdeg$; see \citealt{SurveyOps.Schlafly.2023}). Meanwhile, the backup program footprint covers a wider declination range (roughly $-28\fdg5 < \mathrm{Dec} < 80\arcdeg$) and lower Galactic latitude (roughly $|b| >7\arcdeg$), as backup-program targets are selected from Gaia photometry \citep{gaia-collaboration16a} which is available over the whole sky \citep{TS.Pipeline.Myers.2023, dey25a}.

\begin{deluxetable*}{lccccc}
\tablecaption{Summary of Observational Programs in DR1}
\label{tab:survey_nights_tiles_exp}
\tablehead{
\colhead{} & \colhead{No. of} & \colhead{No. of} & \colhead{No. of} & \colhead{Effective Time\tablenotemark{a}} & \colhead{Area\tablenotemark{b}} \\
\colhead{Program} & \colhead{Nights} & \colhead{Tiles} & \colhead{Exposures} & \colhead{(h)} & \colhead{(deg$^2$)}
}
\startdata
\multicolumn{6}{c}{Survey Validation} \\
\hline
CMX\tablenotemark{c} & 1 &     1 &      4 &    0.9 & 8 \\
SV1\tablenotemark{d} & 90 & 187 & 1674 & 175.3 & 1082 \\
SV2 & 8 &    37 &     70 &    6.4 & 102 \\
SV3 & 38 &   488 &    710 &  102.9 & 197 \\
\hline
\multicolumn{6}{c}{Special Observations} \\
\hline
Special & 38 &    42 &    148 &   13.7 & 243 \\
\hline
\multicolumn{6}{c}{Main Survey} \\
\hline
Bright  & 262 &  2275 &   2569 &  148.8 & 9739 \\
Dark & 212 &  2744 &   3420 &  782.9 & 9528 \\
Backup  & 92 &   327 &    581 &    6.0 & 2726 \\
\enddata
\tablenotetext{a}{The effective exposure time is the on-sky integration time in reference, or ideal conditions (see \S\ref{sec:instrument} and \citealt{Spectro.Pipeline.Guy.2023} for details).}
\tablenotetext{b}{The area covered by tiles is larger than the true effective area available to targets due to bright star exclusions, focal plane geometry, hardware configuration, and higher priority targets blocking lower priority targets.}
\tablenotetext{c}{DR1 includes a single commissioning (CMX) tile (TILEID=80615, \texttt{SURVEY=cmx} covering M33 which was obtained during the early part of SV.}
\tablenotetext{d}{\texttt{SURVEY=sv1} includes both Target Selection Validation tiles and tiles dedicated to secondary targets \citep{DESI2023b.KP1.EDR}.}
\end{deluxetable*}

Although the tile centers are fixed, the actual assignment of targets to fibers is not defined until we observe a particular location on the sky. At that time, the Merged Target List \citep[MTL;][]{SurveyOps.Schlafly.2023} is used to assign the highest-priority target to each fiber, in order to produce a tile with a fixed fiber-assignment configuration. The primary motivation for this ``on the fly'' design strategy is to be able to identify and increase the priority of $z > 2.1$ quasars from DESI spectra so that they will be repeatedly observed.  At $z = 2.1$, the Ly$\alpha$ forest redshifts into the DESI spectral coverage (see \S\ref{sec:instrument}), making these targets especially valuable probes of large-scale structure at high redshifts.  However, because we do not know in advance which quasars are at $z > 2.1$, we are forced to learn this from the DESI spectra themselves. Once Ly$\alpha$-forest QSOs have been identified, we update the MTL with the results to ensure that future DESI observations will observe these targets whenever possible. The MTL allows us to identify times when observations of targets failed, for example, due to being assigned to a positioner which has stopped functioning. Observations of these targets may then be repeated in the future.

The need to use information about past observations in order to inform future observations leads to the following survey operations mode or sequence:
\begin{enumerate*}
\item Observe tiles in regions of the sky where the MTL is up to date with the latest observations;
\item Analyze those tiles (typically the day after observations have been obtained) to determine redshifts, with particular care taken to identify $z > 2.1$ quasars; and
\item Incorporate the resulting redshifts into the MTL, in order to inform future targeting.
\end{enumerate*}

The DESI spectroscopic pipeline processes each night's observations as they come in, and usually redshifts are available the following morning \citep[see \S\ref{sec:datareduction} and][]{Spectro.Pipeline.Guy.2023}.\footnote{The DESI team informally and fondly refers to this process as ``redshifts by breakfast."}  Following a quality assurance process by members of the survey operations team, the reduced data are incorporated into the MTL, roughly twice per week \citep{SurveyOps.Schlafly.2023}.

Tiles are selected for observation each night automatically by the Next Tile Selector \citep[NTS;][]{SurveyOps.Schlafly.2023}. The basic scheme is to prefer equatorial fields while minimizing slewing and obtaining the minimum airmass possible. The NTS avoids observing areas of the sky where observations have been made that have not been incorporated into the MTL, so that any remaining Ly$\alpha$ quasars in those regions can be identified. The NTS is also responsible for deciding whether to observe bright, dark, or backup tiles. This decision is made on the basis of the current survey speed, which is based on real-time measurements of the transparency, seeing, and sky background from the Exposure Time Calculator \citep[ETC;][]{Expcalc.Kirkby.2024}. The survey speed is expressed as a fraction relative to nominal dark conditions, where the sky background is 21.07~mag~arcsec$^{-2}$ in the $r$-band, the seeing is 1\farcs1, and conditions are photometric. When the survey speed is greater than 0.4 we observe the dark program; when the survey speed is greater than 0.08 but less than 0.4 we observe the bright program; otherwise we observe backup program tiles.

\begin{figure*}
\centering 
\includegraphics[width=0.7\columnwidth]{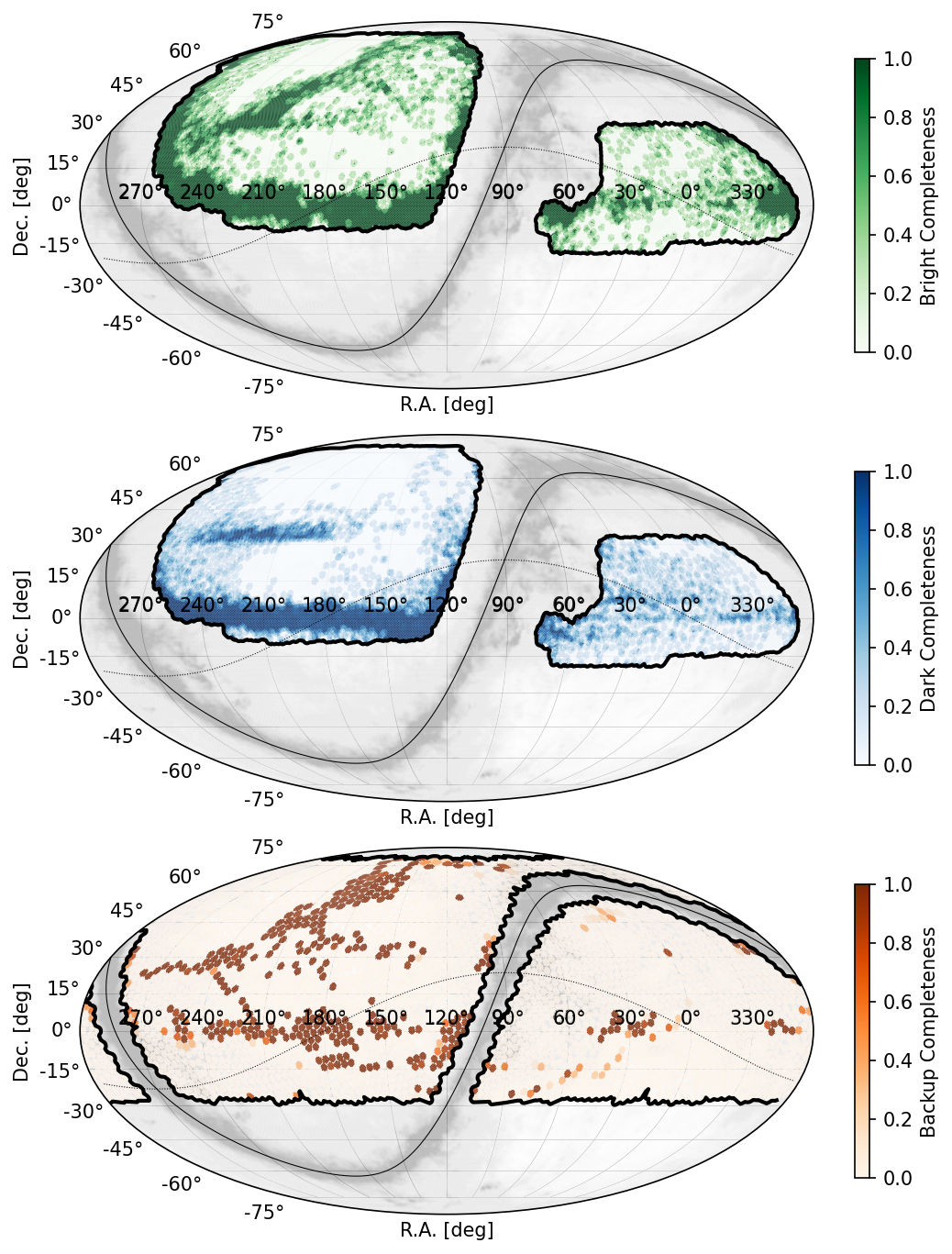}
\caption{Completeness of the DESI main survey based on observations between 2021 May 14 and 2022 June 13 for the bright, dark, and backup programs (from top to bottom). Dark-shaded areas are complete, while white areas have not yet been observed. We use an equal-area Mollweide projection in equatorial coordinates and indicate the per-program DESI main-survey footprint with a thick black curve. As discussed in \S\ref{sec:main}, the bright- and dark-time footprints are identical, while the backup program footprint extends to lower Galactic latitude. The dashed line shows the Galactic plane, the dotted line shows the ecliptic plane, and we also display the Galactic reddening level outside the DESI footprint in gray. 
\label{fig:progress}}
\end{figure*}

We observe tiles for an effective exposure time of 1000~s for the dark program and 180~s for the bright program. Effective exposure times intend to deliver a given ``average'' signal-to-noise ratio on a reference spectrum, and can be thought of as ordinary exposure times in the reference conditions. With these effective exposure times, and accounting for the fact that DESI is not always observing in reference conditions and has overheads and downtime, the DESI instrument can be used to observe a 14,000~deg$^{2}$ survey in five years \citep{SurveyOps.Schlafly.2023}.

The survey had one major shutdown in its first year, between 2021 July 10 and 2021 September 20, when the focal-plane electronics were upgraded. Apart from these interruptions, DESI executed close to continual operation, weather permitting. As shown in Table~\ref{tab:survey_nights_tiles_exp}, more than five thousand main-survey tiles were observed, roughly equally split between the bright and dark programs. The backup program began after the main survey started, on 2021 November 25, and comprises 327 unique observed tiles in DR1.

\begin{figure*}
\centering 
\includegraphics[width=1.01\columnwidth]{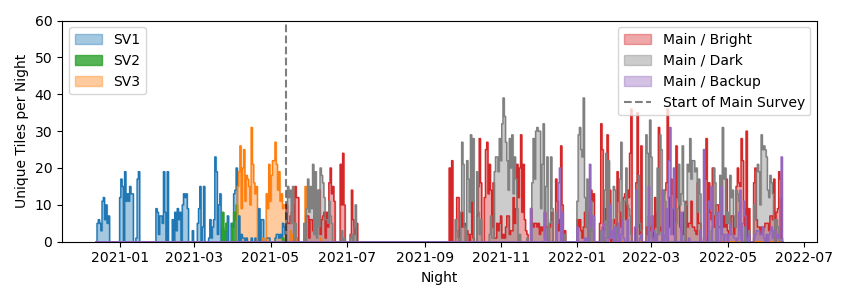}
\caption{Number of unique SV and main survey tiles in DR1 as a function of night (see Table~\ref{tab:survey_nights_tiles_exp}). Note that tiles can be observed on multiple nights and that the vertical dashed line indicates the start of the main survey on 2021 May 14. The large gap in observations between 2021 July 10 and 2021 September 20 was due to a major upgrade of the focal-plane electronics.\label{fig:tiles_per_night}}
\end{figure*}

The data being released in DR1 represents a meaningful fraction of the full DESI survey. Through 2022 June 13, the dark survey was 29.0\% complete, the bright survey was 41.3\% complete; and the backup program was 5.2\% complete. Figure~\ref{fig:progress} shows the spatial distribution of the main-survey completeness on the sky for each of these three programs, and Figure~\ref{fig:tiles_per_night} plots the number of unique tiles observed each night over the window of time spanned by DR1. The start of the main survey on 2021 May 14 is clearly evident as the time when we transitioned from SV3 to the main survey. Though noisy, one can also make out the pattern of the full and new moon, as nightly observations alternate between the dark- and bright-time programs.

In Figure~\ref{fig:progress}, the completeness is strongly affected by the survey strategy and observational conditions.  We aim to observe ``depth first,'' starting at the equator, obtaining all spectra in a particular region of the sky before moving off the equator.  This strategy leads to the equatorial regions with especially dense regions in the bright and dark programs. Kitt Peak occasionally experiences strong southerly winds, which force observation of tiles in the north, leading to the dense regions of observations north of $\mathrm{Dec} = 32\arcdeg$.  These regions are especially prominent in the dark program.  Most bright observations are taken when the Moon is high in the sky, and DESI tries to observe locations more than $50\arcdeg$ away from the Moon, making the distribution of bright tiles less concentrated at the equator than the dark tiles.  Finally, because we need to identify quasars in observations before reobserving any patch of the sky, occasionally we have operated in a ``breadth first'' mode while validating the observations, leading to some large areas of limited coverage in the bright and dark programs.

\subsubsection{Survey Validation}\label{sec:sv}

Although the primary focus of DR1 is on the new main-survey data, DR1 also includes all the SV data which were taken before the start of the main survey:
\begin{enumerate*}
\item SV1 (Target Selection Validation) tiles were used to verify and refine the target-selection algorithms for the main survey, and include dedicated secondary-target tiles which were used for special programs before we developed the special survey (see \S\ref{sec:special}).
\item SV2 (Operations Development) tiles were used to test survey-like DESI operations; and
\item SV3 (One Percent Survey) tiles comprised a high-completeness sample of observations using the final set of DESI target selection algorithms.
\end{enumerate*}

These programs were all part of the EDR and are discussed in extensive detail in \citet{DESI2023b.KP1.EDR}; they remain useful for specialized analyses due to their depth (SV1), high spectroscopic completeness (SV3), and for comparison to past work on the EDR samples.

\subsubsection{Special Tiles}\label{sec:special}

Approximately two percent of DESI observing time in its first year of science operations was spent on the special survey. These tiles cover a variety of use cases, both technical and scientific, and so they are kept separate from the main survey and associated MTL strategy. We summarize these observations in Table~\ref{tab:specialobs} and describe them in more detail in Appendix~\ref{app:tertiarytargets}.

\begin{deluxetable*}{lp{3.5cm}ccp{2.5cm}c}[t]
\tablecaption{Summary of Special Observations in DR1 \label{tab:specialobs}}
\tablehead{
\colhead{} & \colhead{} & \colhead{No. of} & \colhead{No. of} & \colhead{} & \colhead{Effective} \\
\colhead{Observations\tablenotemark{a}} & \colhead{Description} & \colhead{Tiles} & \colhead{Exposures} & \colhead{\texttt{TILEID}s} & \colhead{Time\tablenotemark{b} (h)}
}
\startdata
Bright & Bright test tiles. & 15 &  80 & 80978--80981, 82258--82268 &  2.5 \\
Dark & Dark test tiles; Sgr stream tile. &  4 & 9 & 80977, 81100, 81112, 82237 &  1.2 \\
Backup & Backup test tiles. &  9 & 25 & 82401--82409 &   0.2 \\
M31 & M31 special program. &  2 &   9 & 82634--82635 &  3.3 \\
Odin & LAE/LBG targets in COSMOS. &  1 &   8 & 82636 &  2.7 \\
Tertiary1 & Dense, $z < 21.6$ COSMOS targets. & 11 &  17 & 82637--82647 & 3.9 \\
\enddata
\tablenotetext{a}{All special tiles have \texttt{SURVEY=special}. The bright, dark, and backup special tiles have \texttt{PROGRAM=bright}, \texttt{dark}, and \texttt{backup}, respectively, while the m31, odin, and tertiary tiles all have \texttt{PROGRAM=other}.}
\tablenotetext{b}{Defined in Table~\ref{tab:survey_nights_tiles_exp}.}
\tablecomments{See Appendix~\ref{app:tertiarytargets} for more details regarding these observations.}
\end{deluxetable*}

\section{Data Reduction and Data Products}\label{sec:redux}

In this section we describe the major DR1 data products and how they are produced. First, in \S\ref{sec:datareduction} we present a brief overview of how the two-dimensional spectra are reduced to derive redshifts, redshift-quality flags, and spectral classifications. Next, in \S\ref{sec:sample} we summarize the number of unique redshifts in DR1 and show the sample distributions in redshift and on the sky.
In \S\ref{sec:products} we outline the contents and organization of the data products included in DR1, and in \S\ref{sec:lss} we provide a brief introduction to the large-scale structure (LSS) catalogs used to carry out the DESI cosmological analysis and describe where and how those catalogs can be accessed. Finally, in \S\ref{sec:vac-files} we describe the current set of value-added catalogs (VACs) accompanying DR1.

\subsection{From Raw Spectra to Redshifts}\label{sec:datareduction}


The following three subsections describe the data-reduction pipeline (\S\ref{sec:pipeline}), the determination of redshifts and spectral classifications (\S\ref{sec:redrock}), and the improvements in the spectroscopic pipeline relative to the data released in the EDR (\S\ref{sec:improvements}).

\subsubsection{Spectral Extraction and Calibration}\label{sec:pipeline}

The DESI spectroscopic data-reduction pipeline is described in detail in \citet{Spectro.Pipeline.Guy.2023} and \S3 of \citet{DESI2023b.KP1.EDR}. The primary outputs of this pipeline are sky-subtracted, wavelength-calibrated, and flux-calibrated spectra, including estimated uncertainties, and a \textit{resolution matrix}, which encodes the effective instrument resolution.

The data acquired by DESI at KPNO are automatically transferred to the National Energy Research Scientific Computing Center (NERSC\footnote{\url{https://nersc.gov}}) for reduction, analysis, and archiving. During nominal operations, calibration data at KPNO are acquired during the afternoon preceding a given night of observations. These calibration data include zero exposure-time (bias) frames; dark exposures; arc lamp exposures; and multiple flat-field exposures taken with LED lamps illuminating a white dome screen.

Using these data, we derive a master bias frame for each CCD, the spectral trace coordinates, wavelength calibration parameters, two-dimensional point-spread functions (PSFs), and a flat-field correction for each fiber of each camera. The processing of a scientific exposure acquired during the subsequent night proceeds as follows: We first pre-process each CCD image by subtracting the bias, the overscan level, and the dark current. We mask pixels affected by cosmic-ray hits and CCD defects, convert counts into electrons, and estimate the noise per pixel. We then adjust the coordinates of the spectral traces and the wavelength calibration of each fiber using the known wavelengths of the emission lines from the night-sky spectrum, while offsetting this solution such that the extracted spectra will be in a vacuum Solar System barycentric frame. These coordinate corrections are typically smaller than a CCD pixel relative to the afternoon calibrations. Next, we extract one-dimensional spectra using the \textit{spectroperfectionism} algorithm of \citet{bolton10a}. The results are counts per fiber and wavelength on a common wavelength grid (vacuum, barycentric), along with the resolution matrix, which quantifies the effective instrument resolution of each extracted spectrum as a function of wavelength. The spectra are then flat-fielded and sky-subtracted. Next, we use observations of F-type stars to determine the instrument throughput and to convert the measured counts into calibrated spectral energy density.\footnote{All extracted (one-dimensional) DESI spectra are in units of $10^{-17}~\text{erg}~\text{s}^{-1}~\text{cm}^{-2}~\text{\AA}^{-1}$.} Finally, depending on which data products are being generated (see \S\ref{sec:products}), multiple observations of the target or TILEID (see \S\ref{sec:operations}) are coadded using optimal (inverse variance) weights.

We record the resolution matrix for each spectrum as a band-diagonal sparse matrix. A perfect resolution model should be multiplied by this matrix to achieve the equivalent observed model at the effective resolution of DESI. The underlying mathematics and motivation for the resolution matrix are described in \S3 of \citet{bolton10a} and in \S4.5 and Appendices C and D of \citet{Spectro.Pipeline.Guy.2023}. We refer interested readers to the repository of DESI tutorials (see \S\ref{sec:tutorials}) for worked examples of how to use the resolution matrix in scientific analyses.

For each spectrum, DESI also measures the \textit{template signal-to-noise ratio squared} (\texttt{TSNR2}) for each of its major target classes BGS, LRG, ELG, QSO, and LYA.\footnote{Here, LYA refers to QSOs which have been spectroscopically confirmed to be Ly$\alpha$ forest ($z>2.1$) quasars. Also note that we do not derive a \texttt{TSNR2} value for MWS targets.} These quantities measure a wavelength-averaged, squared, signal-to-noise ratio, where the noise is determined from the current observations (CCD noise, sky level), the signal amplitude depends on the flux calibration, and the variation of the signal with wavelength is designed to best predict the redshift measurement precision of a given target class. For example, \texttt{TSNR2_ELG} gives larger weight to redder wavelengths covering the [\ion{O}{2}] doublet at the redshifts of typical DESI ELG targets, while \texttt{TSNR2_QSO} gives more weight to bluer wavelengths important for measuring QSO emission lines. \texttt{TSNR2_LRG} is scaled to derive the effective exposure time (described in \S\ref{sec:operations}) of DESI dark-time exposures,
while \texttt{TSNR2_BGS} is used to derive the effective exposure time for bright-time tiles, though this is also normalized to the reference dark conditions (see \S4.14 of \citealt{Spectro.Pipeline.Guy.2023} for more details).


\subsubsection{Redshifts and Spectral Classifications}\label{sec:redrock}

With the fully calibrated spectra in-hand, we use Redrock to determine the optimum redshift and spectral classification of each object (\citealt{ross20a, anand24a}, Bailey et al. 2025, in preparation). Briefly, Redrock fits principal component analysis (PCA) templates of three broad, independent classes of objects---stars, galaxies, and quasars, corresponding to spectral type \texttt{STAR}, \texttt{GALAXY}, and \texttt{QSO}, respectively---on a spectral type-dependent grid of redshift. The fit which produces the lowest overall $\chi^{2}$ value yields the best-fitting redshift (\texttt{Z}), redshift uncertainty (\texttt{ZERR}), spectral type (\texttt{SPECTYPE}), and template coefficients (\texttt{COEFF}). In addition, Redrock reports the (minimum) $\chi^2$ of the best fit (\texttt{CHI2}) and $\Delta \chi^2$ (\texttt{DELTACHI2}), which is the difference between the two lowest $\chi^2$ values across all three spectral classes. Interpreted using Gaussian statistics, \texttt{DELTACHI2} represents the statistical significance of the best fit relative to the next best fit; for example, a \texttt{DELTACHI2} value of 25 implies that the best fit is $5\sigma$ ``better'' than the next best fit (not accounting for systematic uncertainties). Finally, besides redshifts and classification, a key metric reported by Redrock is the per-object redshift warning bitmask, \texttt{ZWARN}. In essence, a value of \texttt{ZWARN\text{=}0} indicates that there are no known problems with either the input spectroscopic data nor with the corresponding redshift fit. We refer the reader to the DESI data model\footnote{\url{https://desidatamodel.readthedocs.io/en/latest/bitmasks.html\#zwarn}}, which documents the full set of possible \texttt{ZWARN} bits and their definitions, and to \S3.2 of \citet{DESI2023b.KP1.EDR} for more discussion.

Finally, the DESI pipeline carries out three additional processing steps: emission-line fitting \citep[\texttt{emlinefit};][]{ELG.TS.Raichoor.2023}; a fit to the \ion{Mg}{2}~$\lambda2800$ doublet \citep{QSO.TS.Chaussidon.2023}; and a quasar neural-network classifier called \texttt{QuasarNET} \citep{busca18a, farr20a, green25a}, all three of which are described in more detail in \S3.1.4 of \citet{DESI2023b.KP1.EDR}.\footnote{Colloquially, DESI team members refer to these steps of the pipeline as ``after-burners,'' since they are collaborator-contributed algorithms executed after the original core pipeline Redrock results have been written out.} Detailed analyses show that these after-burners are essential for evaluating successful ELG redshifts and for correctly classifying and measuring the correct redshifts for some classes of QSOs \citep{VIQSO.Alexander.2023, VIGalaxies.Lan.2023, ELG.TS.Raichoor.2023, QSO.TS.Chaussidon.2023, DESI2023a.KP1.SV}, so the after-burner catalogs are included as standard pipeline outputs.


\begin{deluxetable*}{lll}[t]
\tablecaption{Algorithmic Performance of the Spectroscopic Data Pipeline \label{tab:spectro_pipeline_performance}}
\tablehead{Parameter & Value & Reference\tablenotemark{a}}
\startdata
Wavelength precision                & 0.025~\AA & G23 \S4.7.4, Figure 32 \\
Sky Subtraction                     & $<$1\% systematic & G23 \S4.7.5, Figure 34 \\
Spectrophotometric flux calibration & 6\%--10\% & G23 \S4.9.3, Figure 38 \\
Redshift precision -- BGS, ELG      & 10 km s$^{-1}$ & DESI24c \S7.2 \\
Redshift precision -- LRG           & 50 km s$^{-1}$ & DESI24c \S7.2 \\
Redshift precision -- QSO           & 20--125 km s$^{-1}$ ($z\sim0.8-1.8$) & DESI24c \S7.2 \\
Redshift outliers -- BGS, LRG, ELG  & $\leq 0.3\%$ & DESI24c \S7.2 \\
Redshift outliers -- QSO $z<2.1$    & $0.7\%$ & DESI24c \S7.2 \\
Redshift outliers -- QSO $z>2.1$    & $1.8\%$ & DESI24c \S7.2 \\
Radial velocity precision -- MWS  & $\lesssim10$~km~s$^{-1}$, $0.9$~km~s$^{-1}$ systematic & K24 \S5.4.1 \\
\enddata
\tablenotetext{a}{References G23, DESI24c, and K24 are \citet{Spectro.Pipeline.Guy.2023}, \citet{DESI2024.II.KP3}, and \citet{koposov24a}, respectively.}
\end{deluxetable*}

\subsubsection{Pipeline Improvements and Algorithmic Performance}\label{sec:improvements}


Algorithmic updates for the Iron production in DR1 (relative to the Fuji \texttt{specprod} used in the EDR and Guadalupe; see \S\ref{sec:guide}) include new QSO templates split by redshift range \citep{RedrockQSO.Brodzeller.2023}; improved sky subtraction by modeling the measured fiber throughput as a function of positioner location\footnote{\url{https://github.com/desihub/desispec/pull/1801}} which replaces the simpler per-fiber, per-exposure measured normalization described in \S4.7.2 of \citet{Spectro.Pipeline.Guy.2023}; and multiple smaller bug-fixes. In addition, compared to the Fuji release in the EDR, most intermediate pipeline files are now gzipped to save disk space, although the coadded spectra and measured catalog files remain uncompressed for faster read-access (see \S\ref{sec:products}).

\begin{figure}
\centering 
\includegraphics[width=\columnwidth]{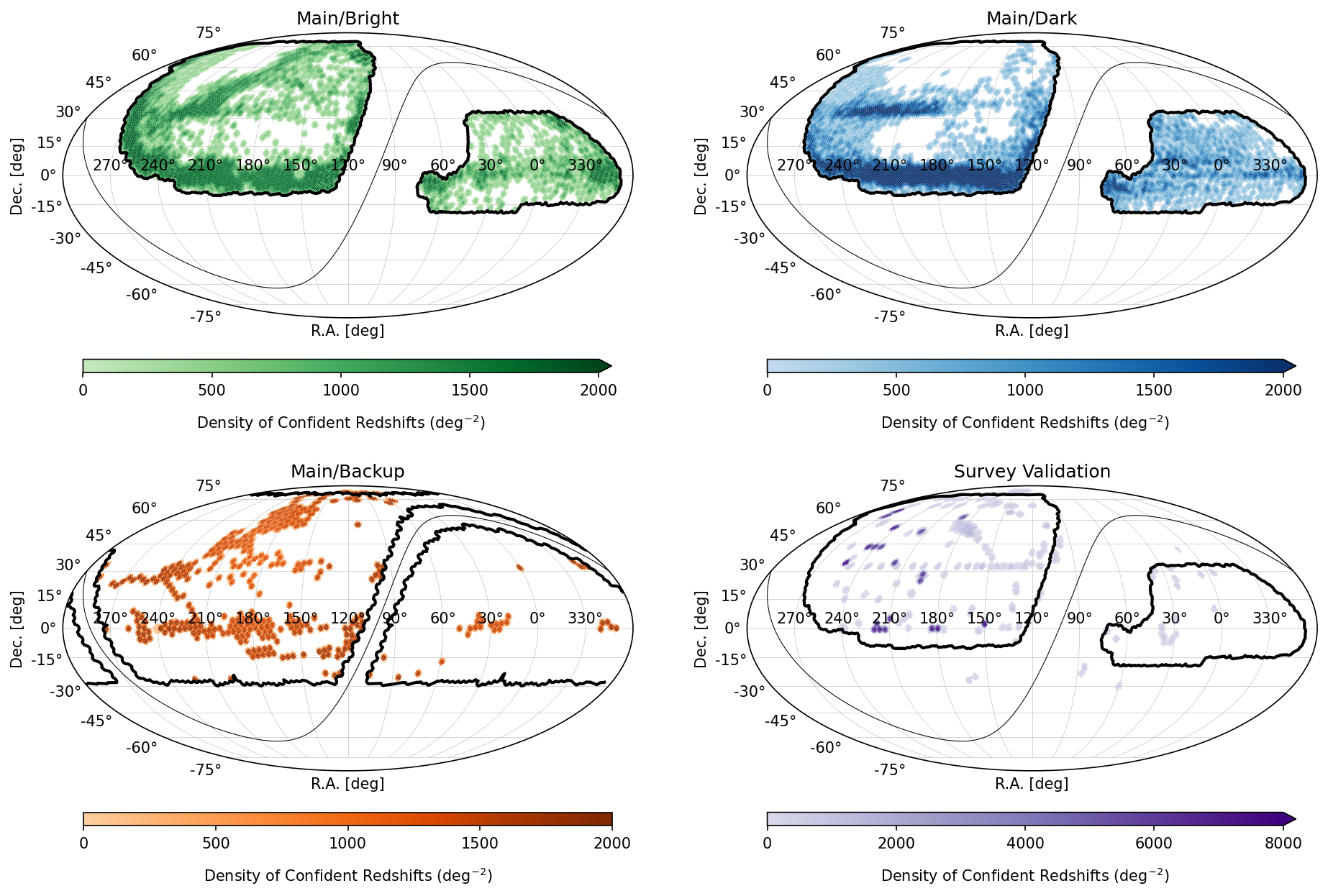}
\caption{Surface density of all unique targets with good redshifts observed in the bright, dark, and backup main-survey programs (first three panels), and in SV (lower-right panel),
rendered using an equal-area Mollweide projection in equatorial coordinates. 
In each panel, the thin gray curve represents the Galactic plane, which divides the DESI footprint into its North and South Galactic Cap regions (shown as thick black outlines; see Figure~\ref{fig:progress}). The density of targets in the bright- and dark-time programs extend to above 2000~deg$^{-2}$, compared to the (as-designed) $>8000$~deg$^{-2}$ surface density of targets in SV.
\label{fig:skymap}} 
\end{figure}  



\begin{figure}
\centering 
\includegraphics[width=1\columnwidth]{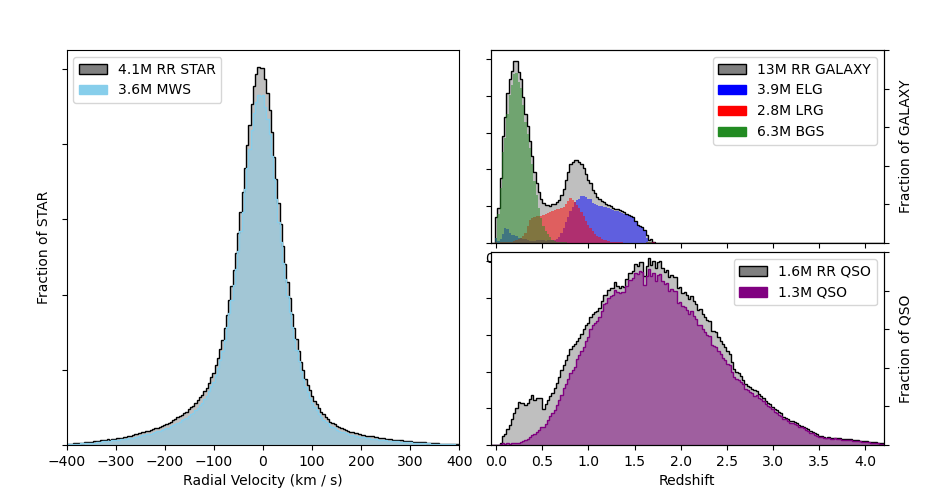}
\caption{Distribution of objects with confidently measured redshifts in DR1 (see Table~\ref{tab:spectra_stats}). In each panel, the gray histograms show objects spectroscopically classified by Redrock as stars (RR \texttt{STAR}), galaxies (RR \texttt{GALAXY}), or quasars (RR \texttt{QSO}) (see \S\ref{sec:redrock}), the light blue histogram shows the radial velocity distribution of MWS targets, and the dark green, red, dark blue, and purple histograms show the redshift distributions of all unique BGS, LRG, ELG, and QSO DESI targets, respectively. Note that a given object can belong to more than one target class and may therefore appear in more than one distribution. \label{fig:nofz-iron}}
\end{figure}  

Table~\ref{tab:spectro_pipeline_performance} provides a summary of the algorithmic performance of the spectroscopic pipeline, including average wavelength precision, sky subtraction residuals, spectrophotometric flux calibration, redshift and radial velocity precision, and redshift outliers. The table includes references to other papers documenting the details.

\subsection{Sample Distributions}\label{sec:sample}

After just 13 months of science operations, DESI has measured confident redshifts for approximately 18.7M unique science targets across all surveys and programs, making DR1 the largest sample of extragalactic redshifts ever assembled. In the main survey alone, DR1 includes a total of 8.5M, 9.0M, and 1.2M objects with reliable redshifts from the bright, dark, and backup programs, respectively. Here and in the rest of this section, we use the (logical ``and'') condition $\texttt{ZCAT\_PRIMARY}\ \text{\&}\ (\texttt{ZWARN}\text{=}0)$ to select a sample of unique objects with confident redshifts. The criterion $\texttt{ZWARN}\text{=}0$ identifies objects with no known hardware, observing, data quality, or redshift-fitting flags (see \S\ref{sec:datareduction}), and \texttt{ZCAT\_PRIMARY} chooses the ``best'' observation of objects which have been observed in different surveys or programs (see \S\ref{sec:products}). For additional details regarding the full range of possible data-quality bit-masks, see the DESI data model.\footnote{\url{https://desidatamodel.readthedocs.io/en/latest/bitmasks.html\#bit-masks-in-desi}} Alternatively, most users of the DESI data will likely be interested in the tracer-dependent quantitative criteria used to define the samples used for DESI cosmological analyses, which are described in \S4.2 of \citet{DESI2023b.KP1.EDR} and \citet{DESI2024.II.KP3} for the EDR and DR1 cosmological results, respectively.

Figure~\ref{fig:skymap} displays the surface density distribution of objects with confidently measured redshifts observed as part of the bright, dark, and backup programs of the main survey (first three panels, starting in the upper-left) and as part of SV (lower-right panel). For reference, the approximate solid angle covered by these four datasets is 9739, 9528, 2726, and 1410~deg$^{2}$, respectively (Table~\ref{tab:survey_nights_tiles_exp}) For more precise tracer-dependent area estimates, see \citealt{KP3s15-Ross}.

\begin{deluxetable*}{cccccccc}[t]
\tablecaption{Number of Confident, Unique Redshifts in DR1\tablenotemark{a} \label{tab:spectra_stats}}
\tablehead{
\colhead{Program} & \colhead{$\mathrm{N}_{\mathrm{MWS}}$\tablenotemark{b}} & \colhead{$\mathrm{N}_{\mathrm{BGS}}$} & \colhead{$\mathrm{N}_{\mathrm{LRG}}$} & \colhead{$\mathrm{N}_{\mathrm{ELG}}$} & \colhead{$\mathrm{N}_{\mathrm{QSO}}$} & \colhead{$\mathrm{N}_{\mathrm{SCND}}$} & \colhead{Total\tablenotemark{c}}
}
\startdata
\multicolumn{8}{c}{Survey Validation} \\
\hline
CMX & 469 & 247 & 1,040 & 734 & 292 & 0 & 3,173 \\
SV1 & 159,115 & 128,921 & 62,706 & 109,420 & 29,387 & 59,538 & 527,927 \\
SV2 & 8,954 & 37,456 & 21,053 & 11,732 & 11,364 & 0 & 89,925 \\
SV3 & 280,253 & 219,213 & 127,876 & 295,831 & 32,843 & 69,811 & 992,821 \\
Total & 448,791 & 385,837 & 212,675 & 417,717 & 73,886 & 129,349 & 1,613,846 \\
\hline
\multicolumn{8}{c}{Special Observations} \\
\hline
Special & 42,690 & 31,303 & 3,778 & 4,407 & 2,599 & 58,252 & 141,473 \\
\hline
\multicolumn{8}{c}{Main Survey} \\
\hline
Backup & 1,192,150 & 0 & 0 & 0 & 0 & 5 & 1,212,427 \\
Bright & 2,237,995 & 5,940,739 & 189,759 & 736 & 4,607 & 705,730 & 8,484,481 \\
Dark & 211,235 & 339,459 & 2,639,852 & 3,925,609 & 1,335,505 & 1,289,489 & 8,962,896 \\
Total & 3,641,380 & 6,280,198 & 2,829,611 & 3,926,345 & 1,340,112 & 1,995,224 & 18,659,804 \\
\hline
\enddata
\tablenotetext{a}{As documented in \S\ref{sec:sample}, we use the criteria $\texttt{ZCAT\_PRIMARY}\ \text{\&}\ (\texttt{ZWARN}\text{==}0)$ to select the sample of unique objects with confident redshifts.}
\tablenotetext{b}{The number of MWS objects listed in this column includes the standard stars used for spectrophotometric calibration.}
\tablenotetext{c}{The total may be different than the sum of the the columns because some targets may belong to one or more target class (or a target class may not be listed).}
\end{deluxetable*}


In Figure~\ref{fig:nofz-iron} we show the radial velocity distribution of stellar targets (left-hand panel) and the redshift distribution of extragalactic targets (right-hand panels) in DR1 with well-measured redshifts. In each panel, the gray histogram shows the distribution of objects spectroscopically classified by Redrock (RR) into stars, galaxies, and quasars (\texttt{SPECTYPE=STAR}, \texttt{GALAXY}, and \texttt{QSO}; see \S\ref{sec:redrock}). The light-blue histrogram shows the radial velocity distribution of MWS targets, and the dark-green, red, dark-blue, and purple histograms show the redshift distribution of BGS, LRG, ELG, and QSO tracers, respectively. As discussed in \citet{DESI2023a.KP1.SV, DESI2023b.KP1.EDR} and the collection of papers listed in Table~\ref{tab:tspapers}, these redshift distributions match the finely-tuned target-selection algorithms developed for each class of DESI target.


\begin{figure*}[t]
\centering 
\includegraphics[width=1\columnwidth]{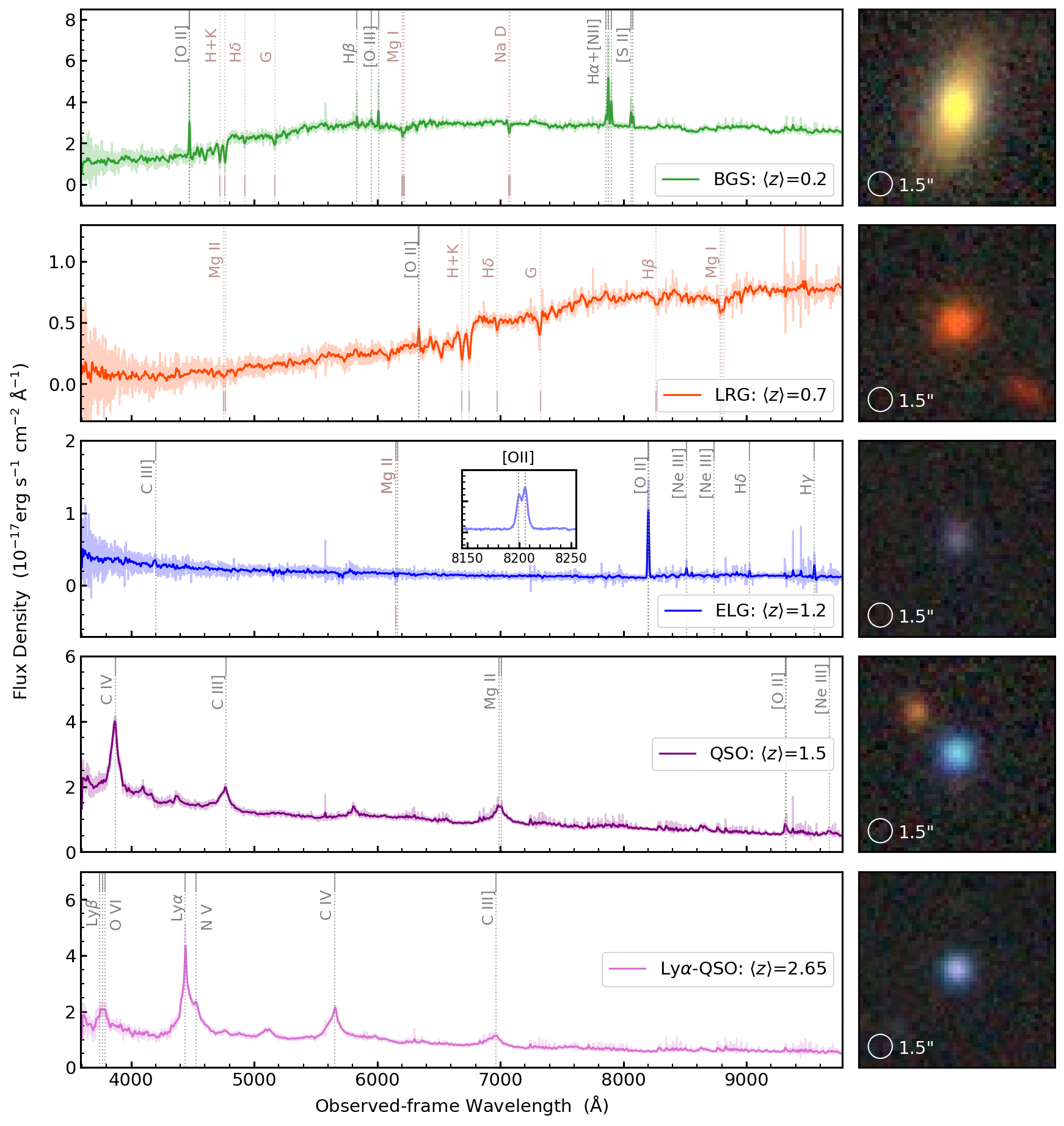}
\caption{Composite spectra and example LS images of the extragalactic target classes. The left-hand panel of each row shows the average of 75-100 spectra in a narrow redshift slice around the mean redshift value labeled in the lower-right of each panel. For the BGS, LRG, and ELG target classes we use a bin width of $\Delta z/(1+z)=10^{-5}$ and for the QSO and Ly$\alpha$ QSO target classes we use $\Delta z/(1+z)=3\times10^{-5}$. The faint colored lines represent the inverse-variance weighted average spectra, while the darker lines are the same spectra smoothed with a five-pixel Gaussian kernel. The vertical dotted lines indicate the expected wavelengths of key emission and absorption lines, and for the ELG class the inset shows the resolved [\ion{O}{2}]~$\lambda\lambda3726,29$ doublet. The right-hand panel of each row features a $g,r,z$ color image of one representative object from each target class with the fiber diameter of 1.5~arcsec drawn to scale. \label{fig:spectra}}
\end{figure*}  

Table~\ref{tab:spectra_stats} summarizes the exact number of unique, well-measured redshifts in DR1. In this table we show the detailed breakdown of the number of good redshifts as a function of target class (including secondary, SCND, targets; see \S\S \ref{sec:firstyear} and \ref{sec:targets}), and the full set of observational programs included in DR1 (see \S\ref{sec:operations}). In addition, the last column gives the total number of good redshifts within each program, although we emphasize that this total cannot be derived from the sum of the preceding columns because it includes objects which may belong to one or more (or none) of the listed target classes.

Finally, in Figure~\ref{fig:spectra} we show examples of coadded spectra and Legacy Surveys color images for each of the primary extragalactic target classes. BGS galaxies tend to include both obvious stellar continuum and emission lines in their spectra with a spatially resolved morphology. LRGs tend to have spectra dominated by a red stellar continuum indicative of old stellar populations and images characterized by red colors with a compact (e.g., de Vaucouleurs) morphology. ELGs have, on-average a fainter and bluer continuum with respect to the other galaxy classes, with distinct [\ion{O}{2}]~$\lambda\lambda3726,29$ doublet emission and their images often appear faint, blue and barely resolved. Finally, QSO spectra are characterized by a blue continuum and broad emission lines arising from the nuclear accretion disk and surrounding ionized gas, and they all have a point-source morphology by selection.

\subsection{Data Products}\label{sec:products}

In this section we provide a high-level overview of the organizational (directory) structure of the data included in DR1 (see \S\ref{sec:access} for information regarding how the data can be accessed). For a complete description of all the directories and files in DR1, as well as their provenance and inter-dependencies, please see the DESI data model.\footnote{\url{https://desidatamodel.readthedocs.io}}

Table~\ref{tab:directory_structure} shows key elements of the DR1 file and directory structure and relative locations, and the number of files and total size in terabytes. Most users will be interested in the data and catalogs in \texttt{spectro/redux/iron/}\footnote{As discussed in \S\ref{sec:guide}, Iron is the primary spectroscopic production for DR1.}, the LSS catalogs in \texttt{survey/catalogs/dr1/LSS/} (see \S\ref{sec:lss}), and the VACs in \texttt{vac/dr1/} (see \S\ref{sec:vac-files} and Appendix~\ref{app:vacs}). 

In the top-level production directory, \texttt{spectro/redux/iron/tiles-iron.fits} contains a catalog of all DESI tiles included in Iron. This file can be used to quickly assess the observational footprint of DR1, and to filter the set of tiles by survey (\texttt{SURVEY}) or program (\texttt{PROGRAM}). Because tiles may be observed on multiple exposures spanning multiple nights (see \S\ref{sec:operations}), more detailed per-exposure information can be found in the \texttt{spectro/redux/iron/exposures-iron.fits} file, for example, for time-domain studies or for comparisons of systematic differences of data obtained on different nights.

In \S\ref{sec:spectra-files} we describe the organization of the spectra, coadds, and lower-level (per-observation) redshift files, and in \S\ref{sec:redshift-catalogs} we describe the merged redshift catalogs, which are joined across different combinations of DESI surveys and programs. In addition, \S\ref{sec:target-catalogs} describes specific updates to the parent target catalogs used for DR1 observations. For additional information and descriptions of the other kinds of files available in DR1, such as the fiber assignment and parent photometric-target catalogs, please see \citet{DESI2023b.KP1.EDR} and the DESI data model.

Finally, Appendix~\ref{app:issues} documents known problems and other caveats regarding the data released in DR1.

\begin{deluxetable*}{lccl}[t]
\tablecaption{
DR1 Directory Structure and Data Volume
}
\label{tab:directory_structure}
\tablehead{
\colhead{} & \colhead{Size} & \colhead{No. of} & \colhead{} \\
\colhead{Directory} & \colhead{(TB)} & \colhead{Files} & \colhead{Description}
}
\startdata
\texttt{spectro/}                         & 266 & 5,921,713 & All spectroscopic data\\
\hspace{0.4cm}\texttt{data/}              & 23 & 234,530 & Raw data \\
\hspace{0.4cm}\texttt{redux/}       & & & Reduced data \\
\hspace{0.8cm}\texttt{iron/}        & 212 & 4,907,360 & Iron spectroscopic production \\
\hspace{1.2cm}\texttt{tiles-iron.fits}  & $<10^{-3}$ & 1 & Unique list of tiles \\
\hspace{1.2cm}\texttt{exposures-iron.fits} & $<10^{-3}$ & 1 & Unique list of exposures \\
\hspace{1.2cm}\texttt{exposures/}   & 61 & 2,435,160 & Intermediate processing files per exposure \\
\hspace{1.2cm}\texttt{healpix/}     & 34 & 565,037 & Spectra and redshifts grouped by HEALPix \\
\hspace{1.2cm}\texttt{tiles/}       & & & Spectra and redshifts grouped by TILEID \\
\hspace{1.6cm}\texttt{cumulative/}  & 35 & 909,161 & Spectra and redshifts coadded across all nights \\
\hspace{1.6cm}\texttt{pernight/}    & 7.9 & 155,299 & Spectra and redshifts coadded within a night \\
\hspace{1.2cm}\texttt{zcatalog/}    & & & \\
\hspace{1.6cm}\texttt{v1/}          & 0.103 & 51 & Merged redshift catalogs \\
\hspace{0.8cm}\texttt{guadalupe/}   & 31 & 769,219 & Like \texttt{iron/} but for the Guadalupe production \\
\texttt{survey/}                    & & & Survey operations and LSS catalog files \\
\hspace{0.4cm}\texttt{catalogs/}   & & & \\
\hspace{0.8cm}\texttt{dr1/}        & & & \\
\hspace{1.2cm}\texttt{LSS/}        & 4.4 & 1,280 & Large-scale structure catalogs \\
\hspace{1.2cm}\texttt{QSO/}        & $<10^{-3}$ & 2 & QSO catalogs \\
\texttt{target/}                    & & & \\
\hspace{0.4cm}\texttt{catalogs/}    & 21 & 61,614 & Input target catalogs \\
\hspace{0.4cm}\texttt{fiberassign/} & 0.062 & 29,020 & Fiber assignment catalogs for each TILEID \\
\texttt{vac/}                       &  &  & \\
\hspace{0.4cm}\texttt{dr1/}         & 11 & 824,203 & Contributed value-added catalogs \\
\enddata
\tablecomments{Please see the DESI data model documentation at \url{https://desidatamodel.readthedocs.io} for more details regarding the directory and subdirectory structure listed in this table, individual file formats, and additional directories and files not listed here used by the DESI pipeline, including calibration files.}
\end{deluxetable*}

\subsubsection{Spectra, Coadds, and Redshifts}\label{sec:spectra-files}

Per-exposure spectra, coadded spectra (coadds), redshifts, and after-burner catalogs are organized into two broad groups or categories: per-tile and ``healpix". Tile-based spectra can be found in \texttt{spectro/redux/iron/tiles/}; these spectra combine information across multiple exposures of the same tile, but not across different tiles (even if the same target was observed on multiple tiles; see \S\ref{sec:guide}). Meanwhile, healpix coadds combine all the available exposures of targets observed on different tiles into a given HEALPix pixel on the sky \citep[nested scheme, using \texttt{NSide=64};][]{gorski05a, zonca19a}; healpix coadds can be found under \texttt{spectro/redux/iron/healpix/}. 

In the case of both the tile-based and healpix coadds, data are not combined across surveys (SV1, SV2, SV3, main, special) or programs (bright, dark, backup). This decision is driven by the desire to prioritize the  uniformity of the data in each survey/program combination (but see the \texttt{ZCAT\_PRIMARY} flag described in \S\ref{sec:redshift-catalogs}). In general, we expect most users to utilize the HEALPix-grouped spectra and redshifts, while the tile-based results will be used for more specialized (including cosmological) analyses.

We organize healpix-grouped outputs in subdirectories by HEALPix number; however, in order to avoid having tens of thousands of subdirectories at the same level, these subdirectories are additionally grouped by integers given by \texttt{int(healpix/100)}. For example, all \texttt{SURVEY=main}, \texttt{PROGRAM=dark} data for targets in nested \texttt{NSide=64} HEALPix number 31542 can be found in \texttt{spectro/redux/iron/healpix/main/dark/315/31542/}. The files in this directory include the per-exposure (uncoadded) spectra, coadded spectra (across exposures but not across cameras), Redrock redshift fits and classifications, and the outputs from the pipeline after-burner algorithms (described in \S\ref{sec:redrock}).

Meanwhile, tile-based spectra, coadds, redshifts, and after-burner catalogs contain the same set of files, but are organized differently in additional subgroups under \texttt{spectro/redux/iron/tiles/} (see Table~\ref{tab:directory_structure}). The \texttt{cumulative/} directory tree contains all data for each tile, coadded across exposures and nights, while the \texttt{pernight/} directory tree combines data within a night but not across nights, enabling reproducibility studies of the same targets observed under different conditions on different nights.\footnote{For the Iron spectroscopic production,  the \texttt{pernight} grouping was only created for SV tiles, not main-survey tiles.} Finally, if a tile was only observed on a single night, the contents of the \texttt{cumulative/} and \texttt{pernight/} directories are effectively identical, but they are still kept in both directories so that each can be used independently.


\subsubsection{Merged Redshift Catalogs}\label{sec:redshift-catalogs}

Merged redshift catalogs, combined across thousands of smaller, individual files and catalogs, can be found in \texttt{spectro/redux/iron/zcatalog/v1/}. Like the spectra and coadds described in \S\ref{sec:spectra-files}, these catalogs come in multiple groups, for example, combining all the cumulative, tile-based redshifts for a given survey and program into a single file, with different files for different survey/program combinations. Future data releases may include \texttt{v2} (or higher) version numbers with a modified data model optimized for the increasingly large catalogs. In general, we recommend using the highest version number available, which is \texttt{v1} in the case of DR1.

For analyses seeking the ``best'' redshift for a given target, regardless of the DESI-specific survey/program combination, we recommend the \texttt{zall-pix-iron.fits} file, which combines all the HEALPix-based redshifts across all surveys and programs into a single file. The ``best" redshift can be retrieved using the \texttt{ZCAT\_PRIMARY} boolean column, which is derived using the Python script \texttt{desispec.zcatalog.find_primary_spectra}.\footnote{\url{https://github.com/desihub/desispec/blob/0.60.2/py/desispec/zcatalog.py\#L58}} This same script can be adopted to subselect multiply-observed targets from a custom selection of spectra to determine the recommended redshift. Similarly, the \texttt{zall-tilecumulative-iron.fits} file provides all cumulative, tile-based redshifts across all surveys and programs, with \texttt{ZCAT\_PRIMARY} indicating the recommended best single tile-based redshift per target.

\subsubsection{Parent Target Catalogs}\label{sec:target-catalogs}

Target catalogs used as input for DESI observations were previously published in the Early Target Selection (ETS) release described in \citet{TS.Pipeline.Myers.2023} and included as part of the EDR.\footnote{\url{https://data.desi.lbl.gov/public/edr/target}} The photometric catalogs from which \textit{primary} DESI targets have been selected remain unchanged relative to the EDR and can be found in the \texttt{target/catalogs} directory tree (see Table~\ref{tab:directory_structure}). Meanwhile, some catalogs of secondary\footnote{\url{https://data.desi.lbl.gov/public/dr1/target/catalogs/dr9/1.3.0}} and calibration\footnote{\url{https://data.desi.lbl.gov/public/dr1/target/catalogs/dr9/2.2.0}} targets have been updated between the ETS and DR1 in a manner consistent with the schemas described in \S3.2, \S4.3 and \S4.4 of \citet{TS.Pipeline.Myers.2023}. Additional details regarding secondary targets can be found in Appendix~\ref{app:secondarytargets}; in addition, some random catalogs \citep[see \S 4.5.1 of][] {TS.Pipeline.Myers.2023} were repackaged into HEALPix \citep{gorski05a} pixels to simplify LSS analyses, and these are also included as part of DR1.\footnote{\url{https://data.desi.lbl.gov/public/dr1/target/catalogs/dr9/2.4.0}}

\subsection{Large-Scale Structure Catalogs}\label{sec:lss}

Together with the raw and intermediate DESI data, DR1 includes all the LSS catalogs needed to reproduce the DR1 clustering and cosmology analyses, including the BAO measurements \citep{DESI2024.III.KP4}, the full-shape analyses \citep{DESI2024.VII.KP7B}, and the derivation of the baseline cosmological constraints from the four extragalactic DESI tracers \citep[BGS, LRG, ELG, and QSO;][]{DESI2024.VI.KP7A}. For complete details regarding the sample definitions and characteristics of the LSS clustering catalogs in DR1, see \citet{DESI2024.II.KP3} and \citet{KP3s15-Ross}. Here, we briefly summarize the broad organizational structure of the LSS catalogs, and how they can be accessed (see also Table~\ref{tab:directory_structure}).

\begin{deluxetable}{lp{7cm}p{6cm}c}[t]
\tablecaption{DR1 LSS Catalogs\label{tab:lss_versions}}
\tablehead{
\colhead{Version\tablenotemark{a}} &  \colhead{Description} & \colhead{References} & \colhead{Software Tag}
}
\startdata
v1.2 & Baseline used in DR1 BAO analysis and cosmological results. & \citet{DESI2024.III.KP4, DESI2024.VI.KP7A} & \href{https://github.com/desihub/LSS/releases/tag/v1.2-DR1}{v1.2-DR1} \\
v1.5 & Minor bug fixes; should be used for all clustering measurements at scales greater than the fiber patrol radius ($\gtrsim 90^{\prime\prime}$.) & \citet{DESI2024.VII.KP7B, ChaussidonY1fnl} & \href{https://github.com/desihub/LSS/releases/tag/v1.5-DR1}{v1.5-DR1} \\
\enddata
\tablenotetext{a}{Version of the LSS pipeline used to generate the catalogs.}
\end{deluxetable}

Unlike in the EDR, where the LSS catalogs were released as a VAC (see \S4 of \citealt{DESI2023b.KP1.EDR}), the LSS catalogs in DR1 can be found under the \texttt{survey/catalogs/dr1/LSS} directory (see \S\ref{sec:products} and Table~\ref{tab:directory_structure}). This directory contains all the auxiliary files defining the DR1 tiles, potential assignments for 18 random catalogs, and the input redshifts from the Iron spectroscopic production.

The ready-to-use clustering catalogs and associated randoms are available in \texttt{survey/catalogs/dr1/LSS/iron/LSScats}. In this directory, there are two distinct versions of these catalogs, \texttt{v1.2} and \texttt{v1.5} (see Table 9). A discussion of the differences between these versions can be found in Appendix~B of \citet{DESI2024.II.KP3}.


In addition, we publish all the initial and intermediate data-products needed to process the DESI data with the DESI LSS pipeline.\footnote{\url{https://github.com/desihub/LSS}}
We list the specific tag of the pipeline used to produce each version of the LSS catalogs in the last column of Table~\ref{tab:lss_versions}. The LSS pipeline and these software tags are described in \citet{KP3s15-Ross}, while the content and column descriptions of all the LSS files can be found at the DESI data model documentation.\footnote{\url{https://desidatamodel.readthedocs.io}}

Data containing the information on 128 alternative realizations of the DR1 fiber-assignment history (used, e.g., to determine pairwise inverse probability, PIP, weights for 2-point clustering measurements; see \citealt{KP3s7-Lasker} and \citealt{KP3s6-Bianchi}), together with the catalog version \texttt{v1.5pip} described in Appendix~C of \citet{DESI2024.II.KP3}, will be published in the near future.

Finally, dedicated mocks to assess the quality of the data, the completeness weights, and to derive covariance matrices, will be published at a later time using a similar directory structure and data model.

\subsection{Value-Added Catalogs}\label{sec:vac-files}

Each DESI data release includes multiple VACs, which are additional data products, catalogs, and documentation contributed by members of the DESI science collaboration. These VACs are built upon the core data products (spectra, classifications, redshifts) from this data release, and include additional data useful for a variety of scientific analyses. DR1 includes a total of 27 distinct VACs at the date of the public data release, although additional VACs may be added to DR1 following this date. The most up-to-date list of VACs and their associated documentation and references can be found online at the public DESI website.\footnote{\url{https://data.desi.lbl.gov/doc/vac}} 

Table \ref{tab:vacs} provides a brief overview of the VACs released concurrently with DR1, along with short descriptions of the contents (and associated reference, if available) of each VAC. Additional details regarding each VAC can be found in Appendix~\ref{app:vacs}.

\begin{singlespace}

\startlongtable
\begin{deluxetable*}{l @{\hskip 14pt} p{0.72\textwidth}}
\tablewidth{0pt}
\tablecaption{Summary of VACs in DR1.\label{tab:vacs}}
\tablehead{\colhead{VAC Name} & \colhead{Description}}
\startdata
\multicolumn{2}{c}{General VACs} \\
\hline
LS/DR9 Photometry & Merged targeting catalogs and Legacy Surveys DR9 Tractor photometric
  catalogs for all observed and potential DESI targets. \\
Sky Spectra & Example sky spectra with detailed metadata from the DESI pipeline. \\
BAO Cosmology Results & Cosmology chains and posterior maximization results for the DESI DR1 BAO cosmology results \citep{DESI2024.VI.KP7A}. \\
Full Shape Cosmology Results & Cosmology chains and posterior maximization results for the DESI DR1 full shape analysis results \citep{DESI2024.VII.KP7B}. \\
\hline
\multicolumn{2}{c}{Milky Way Survey (MWS)} \\
\hline
MWS & Analysis of stellar spectra by the MWS Working Group \citep{koposov25a}. \\
MWS BHB & Catalog of spectroscopically confirmed blue horizontal branch (BHB) stars
  \citep{bystrom25a}. \\
MWS SpecDis & Spectrophotometric distances for $\approx4$ million stars in DR1 predicted using a
  neural network trained on stellar spectra \citep{li25a}. \\
SPDist & Spectrophotometric distances for all stars observed by the MWS predicted using a
  multi-layer perceptron trained on a selection of stellar parameters. \\
Stellar Reddening & Spectra and catalog of stars used in dust reddening measurements \citep{zhou24a}. \\
\hline
\multicolumn{2}{c}{Extragalactic Science} \\
\hline
DESI HETDEX & HETDEX and DESI spectra for Hobby-Eberly Telescope Dark Energy Experiment
  (HETDEX) Ly$\alpha$ emitter candidates observed by DESI \citep{landriau25a}. \\
DESIVAST & Cosmic voids identified within the DESI DR1 volume \citep{rincon25a}. \\
Dwarf Galaxy & Extragalactic dwarf galaxies identified in DESI DR1. \\
EmFit & Emission-line fitting results for $z\leq0.45$ galaxies \citep{pucha25a}. \\
FastSpecFit & Spectrophotometric fitting results from the FastSpecFit stellar continuum and
  emission-line modeling code. \\
Extended Halo-based Group & Halo-based group catalog based on Legacy Surveys DR9 for $z$-band apparent
  magnitude $z<21$ galaxies \citep{yang21a}. \\
Mass EMLines & Stellar mass and emission line measurements for galaxies in DR1 \citep{zou24b}. \\
Strong Lensing & Catalog of spectroscopic observations of strong lenses observed in DESI DR1. \\
\hline
\multicolumn{2}{c}{Quasar Science} \\
\hline
AGN/Galaxy Classification & AGN and QSO identifications for galaxies from all target classes in DESI DR1 (Juneau et~al. 2026, in preparation). \\
AGN Host Properties & Stellar masses and other physical properties from spectral energy distribution
  modeling which includes AGN templates \citep{siudek24a}. \\
BHMass & Iron-corrected supermassive black hole masses based on \ion{Mg}{2} at $0.6<z<1.6$
  \citep{pan25a}. \\
\ion{C}{4} Absorbers & Catalog of \ion{C}{4} absorber systems in DESI quasars \citep{anand25a}. \\
DLA NN and GP Finder & DLA parameters and detections using the NN and GP DLA finders
  \citep{wang22a, ho20a}. \\
DLA Template Finder & DLA parameters and detections using DLA Toolkit \citep{brodzeller25a}. \\
MgII Absorber & Summarized information of \ion{Mg}{2} absorption systems in DESI quasars
  \citep{napolitano23a}. \\
ZLyA & Updated redshifts and BAL information used in the Ly$\alpha$ Y1 BAO analysis
  \citep{DESI2024.IV.KP6}. \\
\hline
\multicolumn{2}{c}{Ly$\alpha$ Forest} \\
\hline
Ly$\alpha$ Forest Y1 Deltas & Measured flux-transmission field used in the Ly$\alpha$ Y1 BAO analysis
  \citep{DESI2024.IV.KP6}. \\
Ly$\alpha$ Forest Y1 Correlations & Measured correlations, distortion matrices and covariances used for the
  Ly$\alpha$ Y1 BAO analysis \citep{DESI2024.IV.KP6}. \\
\enddata
\end{deluxetable*}

\end{singlespace}

\section{Data Access}\label{sec:access}


In this section we provide a brief overview of how to access DR1. At the time of this writing, DESI catalogs and spectra can be accessed via file download (\S\ref{sec:files}) or a searchable database (\S\ref{sec:database}). We demonstrate these access methods, as well as techniques for exploring and manipulating DESI data, through a variety of tutorials and examples (\S\ref{sec:tutorials}). The DESI collaboration may deploy additional data-access methods in the future, such as web services designed for interactive data exploration or interfaces enabling users to download individual spectra or custom collections of spectra. The latest information on how to access DESI data can always be found in the DESI documentation.\footnote{\url{https://data.desi.lbl.gov/doc/access}}

All DESI data are released under the Creative Commons Attribution 4.0 International License.\footnote{\url{https://creativecommons.org/licenses/by/4.0}} This license allows users to share, copy, redistribute, adapt, transform, and build upon the DESI data for any purpose, including commercially, as long as attribution is given by citing this paper and including the complete text of the DESI acknowledgment.\footnote{\url{https://data.desi.lbl.gov/doc/acknowledgments}}

\subsection{File Access}\label{sec:files}

The DR1 data directory tree and individual files can be explored and downloaded directly from \url{https://data.desi.lbl.gov/public/dr1}, following the basic structure illustrated in Table~\ref{tab:directory_structure}. 
For efficient bulk-download of files, we provide (and recommend) the Globus\footnote{\url{https://globus.org}} endpoint called ``DESI Public Data.'' Finally, for individuals with access to NERSC, e.g., through other U.S. Department of Energy-sponsored programs, the same files are directly available without any restriction relative to the following top-level directory: \texttt{\$CFS\_DIR/desi/public/dr1/}.

The complete DR1 dataset occupies more than 280~TB of storage. To optimize the use of resources, we strongly recommend that users exercise discretion when downloading data and select only the data essential for their specific analysis. 

We anticipate that the majority of users will initiate their data exploration with one of the merged redshift catalogs, \texttt{spectro/redux/iron/zcatalog/v1/} (containing all objects) or \texttt{survey/catalogs/dr1/LSS/} (containing objects used in DESI cosmological anlyses), as described in \S\ref{sec:redshift-catalogs} and \S\ref{sec:lss}, respectively. Subsequently, users can further refine their data selection by focusing on the objects of particular interest. Once the selection has been defined, users can proceed to download only the files containing spectra relevant to their analysis (see \S\ref{sec:tutorials}).

\subsection{Database Access}\label{sec:database}

When available, databases can provide users with a more flexible and light-weight approach to search vast amounts of data compared to downloading large files. For convenience, catalog-level data---containing target photometry, fiber-assignments, exposure metadata, spectral classifications, and redshift information---are accessible through a searchable PostgreSQL database.\footnote{\url{https://www.postgresql.org}} Detailed information about the structure of the tables and access credentials for NOIRLab’s Astro Data Lab and NERSC can be found in the DESI documentation portal.\footnote{\url{https://data.desi.lbl.gov/doc/access/database}}



NOIRLab’s Astro Data Lab platform offers anonymous public access\footnote{\url{https://datalab.noirlab.edu/desi}} to DESI data via a web-query interface and a Table Access Protocol (TAP) handle.\footnote{TAP-aware clients such as TOPCAT \citep{taylor05a} can point to \url{https://datalab.noirlab.edu/tap} and select the \texttt{desi\_dr1} database.} Additionally, the Astro Data Lab platform offers authenticated access via a JupyterLab\footnote{\url{https://jupyter.org/}} server, as well as access to the full-depth DESI spectra through the SPectra Analysis and Retrievable Catalog Lab (\texttt{SPARCL}\footnote{\url{https://astrosparcl.datalab.noirlab.edu}}), which features a spectral database with a programmatic interface \citep{juneau24a}. The subset of spectra available via SPARCL is limited to the 18.7M healpix-coadded spectra (see \S\ref{sec:spectra-files}) that have been combined across cameras. Other types of spectra 
and files are available at the file-based archive at NERSC previously described in \S\ref{sec:files}.

Finally, individuals with NERSC access can query the DESI database using SQL, or they can utilize the pre-installed \texttt{specprodDB}\footnote{\url{https://github.com/desihub/specprod-db}} Python code, which provides convenient \texttt{SQLAlchemy} \citep{bayer12a} wrapper objects, indexed using q3c \citep{koposov06a, koposov19a}, for rapid, streamlined data-access and manipulation.

\subsection{Tutorials}\label{sec:tutorials}


The DESI collaboration creates and maintains a variety of tutorials in the form of Jupyter notebooks \citep{kluyver16a, juneau21a}. These tutorials are intended to facilitate the introduction of various data products and methods for accessing them and are arranged into thematic and topical sub-directories. Due to the significant data volume and complexity, these tutorials may be used as an entry point to get started with DESI data access and analysis. The main GitHub repository for DESI tutorials is located at: \url{https://github.com/desihub/tutorials}. 



As additional tutorials become available, the DESI documentation will be updated accordingly.\footnote{\url{https://data.desi.lbl.gov/doc/tutorials/}} 


\section{Summary}\label{sec:summary}

This paper presents DESI DR1, the second major public release of DESI data, following the DESI EDR in 2023 June \citep{DESI2023a.KP1.SV, DESI2023b.KP1.EDR}. DR1 includes all the data obtained by DESI during its first 13 months of science operations (2021 May 14 through 2022 June 13), roughly spanning the first year of its 5-year, 14,000~deg$^{2}$ spectroscopic survey, as well as an improved and uniform reprocessing of all  SV data previously released in the EDR. 

The DR1 main survey includes high-quality redshifts for approximately 18.7M unique objects, of which 13.1M are spectroscopically classified as galaxies, 1.6M are quasars, and 4M are stars. Viewed another way, 5.9M of the objects in DR1 are BGS targets observed over $\approx9700$~deg$^{2}$ in the bright-time program; 2.6M, 3.9M, and 1.3M are LRG, ELG, and QSO targets, respectively, observed over $\approx9500$~deg$^{2}$ in the dark-time program; 3.6M are MWS stars observed as part of the bright-time, dark-time, and backup programs; and the remaining 1.4M objects are secondary, tertiary, and other classes of objects observed as part of a variety of new, bespoke science drivers. By way of comparison, we estimate that DESI DR1 contains high-fidelity redshifts for more unique extragalactic objects than all previous SDSS surveys combined by nearly a factor of four.


We summarize the observations contained in DR1 across all surveys and programs, and include a high-level overview of the DESI target-selection algorithms and a description of how DESI carries out the survey on the timescale of days, months, and years. We estimate that the bright- and dark-time main-survey programs are 41.3\% and 29.0\% complete, respectively, based on the data in DR1. We also describe the basic data-reduction, redshift-fitting, and spectroscopic classification algorithms, and show the distribution of targets with well-measured redshifts in celestial coordinates and in redshift. 

By all metrics, DESI outperformed expectations during its first year of science operations and is well ahead of schedule. By the end of its 5-year survey, we estimate that DESI will have measured precise redshifts for approximately 50M unique galaxies and quasars and 25M stars in the Milky Way Galaxy.

Finally, we document how the data are organized and can be accessed publicly. The data being released in DR1 include not only individual and coadded spectra and redshift catalogs, but also the LSS catalogs used in all the DESI year-one cosmological analyses and 24 VACs spanning a broad range of astrophysical classes of objects and scientific scope.

All the figures and key statistics in this paper have been produced using DR1 files and Jupyter notebooks and Python code in an open-source repository.\footnote{\url{https://github.com/desihub/dr1paper}} The Digital Object Identifier (DOI) for DR1 is \texttt{10.5281/zenodo.15089588}, which includes all the data used to generate the figures and tables in this paper, as well as the figures themselves.\footnote{\url{https://zenodo.org/records/15089588}}

\begin{acknowledgments}
This material is based upon work supported by the U.S. Department of Energy (DOE), Office of Science, Office of High-Energy Physics, under Contract No. DE–AC02–05CH11231, and by the National Energy Research Scientific Computing Center, a DOE Office of Science User Facility under the same contract. Additional support for DESI was provided by the U.S. National Science Foundation (NSF), Division of Astronomical Sciences under Contract No. AST-0950945 to the NSF’s National Optical-Infrared Astronomy Research Laboratory; the Science and Technology Facilities Council of the United Kingdom; the Gordon and Betty Moore Foundation; the Heising-Simons Foundation; the French Alternative Energies and Atomic Energy Commission (CEA); the National Council of Humanities, Science and Technology of Mexico (CONAHCYT); the Ministry of Science, Innovation and Universities of Spain (MICIU/AEI/10.13039/501100011033), and by the DESI Member Institutions: \url{https://www.desi.lbl.gov/collaborating-institutions}.

The DESI Legacy Imaging Surveys consist of three individual and complementary projects: the Dark Energy Camera Legacy Survey (DECaLS), the Beijing-Arizona Sky Survey (BASS), and the Mayall z-band Legacy Survey (MzLS). DECaLS, BASS and MzLS together include data obtained, respectively, at the Blanco telescope, Cerro Tololo Inter-American Observatory, NSF’s NOIRLab; the Bok telescope, Steward Observatory, University of Arizona; and the Mayall telescope, Kitt Peak National Observatory, NOIRLab. NOIRLab is operated by the Association of Universities for Research in Astronomy (AURA) under a cooperative agreement with the National Science Foundation. Pipeline processing and analyses of the data were supported by NOIRLab and the Lawrence Berkeley National Laboratory. Legacy Surveys also uses data products from the Near-Earth Object Wide-field Infrared Survey Explorer (NEOWISE), a project of the Jet Propulsion Laboratory/California Institute of Technology, funded by the National Aeronautics and Space Administration. Legacy Surveys was supported by: the Director, Office of Science, Office of High Energy Physics of the U.S. Department of Energy; the National Energy Research Scientific Computing Center, a DOE Office of Science User Facility; the U.S. National Science Foundation, Division of Astronomical Sciences; the National Astronomical Observatories of China, the Chinese Academy of Sciences and the Chinese National Natural Science Foundation. LBNL is managed by the Regents of the University of California under contract to the U.S. Department of Energy. The complete acknowledgments can be found at \url{https://www.legacysurvey.org/}.

Any opinions, findings, and conclusions or recommendations expressed in this material are those of the author(s) and do not necessarily reflect the views of the U. S. National Science Foundation, the U. S. Department of Energy, or any of the listed funding agencies.

The authors are honored to be permitted to conduct scientific research on I’oligam Du’ag (Kitt Peak), a mountain with particular significance to the Tohono O’odham Nation.
\end{acknowledgments}

\facilities{Mayall (DESI), Mayall (Mosaic-3), Blanco (DECam), Bok (90Prime), WISE, NEOWISE, Gaia, NERSC, Astro Data Lab.}

\software{
Astropy \citep{astropy-collaboration13a, astropy-collaboration18a, astropy-collaboration22a}, 
fitsio (\url{https://github.com/esheldon/fitsio}),
healpy \citep{zonca19a},
Matplotlib \citep{hunter07a},
NumPy \citep{harris20a},
SciPy \citep{virtanen20a},
SPARCL \citep{juneau24a}.
}

\appendix
\restartappendixnumbering

\section{Guadalupe Spectroscopic Production}\label{app:guadalupe}

DR1 includes Guadalupe as a supplementary spectroscopic production, covering 653 tiles observed in the first two months of the DESI main survey from 2021 May 14 through 2021 July 9. It was run at the same time and using the same software tags as the Fuji production of the SV data which was released in the EDR \citep[see Table~\ref{tab:drsummary};][]{DESI2023b.KP1.EDR}. The purpose of Guadalupe was to provide a standardized dataset for early DESI main-survey analyses (see \S\ref{sec:guide}). We include Guadalupe in DR1 in support of those early publications, but in all other regards it is superseded by the Iron spectroscopic production and is not recommended for any new work.


The only code and data model difference with respect to the Fuji data described in \citet{DESI2023b.KP1.EDR} are bug fixes to coadded metadata quantities in the redshift catalogs. The original catalogs, equivalent to Fuji, were moved to \texttt{guadalupe/zcatalog/v0/} and the patched catalogs were created in \texttt{guadalupe/zcatalog/v1/} (see Table~\ref{tab:directory_structure}). See Appendix~\ref{app:issues} for further details regarding these bug fixes, including a list of impacted columns (the redshifts and classifications themselves are unchanged).

For the targets in common between Guadalupe and Iron, redshifts are consistent with a scatter of 1.3~km~s$^{-1}$ and a median offset of $+0.04$~km~s$^{-1}$, with fewer than 1\% of targets changing status from good to bad; consequently, the core results
of the papers that used Guadalupe remain unchanged. 


LSS catalogs were constructed from the main-survey tiles included in Guadalupe. The LSS catalog pipeline was at the stage described in \cite{DESI2023b.KP1.EDR} and \cite{BAO.EDR.Moon.2023} presents the choices specific to the Guadalupe LSS catalogs. Referring to the directory tree in Table~\ref{tab:directory_structure}, the Guadalupe LSS catalogs can be found in \texttt{vac/dr1/lss/guadalupe/v1.0}; the data model of these catalogs generally match the Fuji LSS catalogs which were released in the EDR as a VAC (see \S4.2 of \citealt{DESI2023b.KP1.EDR}).

\section{Primary Targets}\label{app:primarytargets}

This appendix outlines the primary targeting bit names, values, and descriptions for the DESI main survey. For details about earlier phases of DESI, such as SV1 (Target Selection Validation) and SV3 (One-Percent Survey), see the appendix of \citet{DESI2023b.KP1.EDR}. 

Main survey bitmasks are recorded in fibermap columns \texttt{DESI\_TARGET}, \texttt{BGS\_TARGET}, and \texttt{MWS\_TARGET}.\footnote{Bitmasks such as \texttt{SV3\_DESI\_TARGET} \citep{TS.Pipeline.Myers.2023} are also typically present in the fibermap, but will be zero for Main Survey targets. Similarly, \texttt{DESI\_TARGET} will be zero for other survey types.} Table~\ref{table:maindark} lists the \texttt{DESI\_TARGET} bits for dark-time targets and calibration targets such as standard stars and sky locations. Table~\ref{table:mainbgs} lists the \texttt{BGS\_TARGET} bits for the Bright Galaxy Survey. Table~\ref{table:mainmws} lists the \texttt{MWS\_TARGET} bits for Milky Way Survey targets, together with some bits that represent calibration targets derived solely from Gaia instead of from Legacy Surveys imaging. These target selection bits are also defined programmatically in the open-source \texttt{desitarget}\footnote{\url{https://github.com/desihub/desitarget}} software package \citep{desitarget2023}. A YAML-format file describing the bits is in subdirectory \texttt{py/desitarget/data/targetmask.yaml},
with convenience wrapper objects in the Python module \texttt{desitarget.targetmask}.
Examples of accessing these bitmasks using this code can be found in \S2 of \cite{TS.Pipeline.Myers.2023}.

\begin{deluxetable*}{ccc}[htb]
\tablecaption{Dark-time targeting bits for the DESI Main Survey.\label{table:maindark}}
\tablewidth{0pt}
\tablehead{
\colhead{Bit-name} &
\colhead{Bit-value} &
\colhead{Description}
}
\startdata
\texttt{LRG} & 0 & LRG \\
\texttt{ELG} & 1 & ELG \\
\texttt{QSO} & 2 & QSO \\
\texttt{QSO\_HIZ} & 4 & QSO selected using high-redshift Random Forest \\
\texttt{ELG\_LOP} & 5 & ELG at standard (ELG) priority \\
\texttt{ELG\_HIP} & 6 & ELG randomly increased to higher (LRG) priority \\
\texttt{ELG\_VLO} & 7 & Very-low priority ELG (filler) \\
\texttt{LRG\_NORTH} & 8 & LRG cuts tuned for Bok/Mosaic data \\
\texttt{ELG\_NORTH} & 9 & ELG cuts tuned for Bok/Mosaic data \\
\texttt{QSO\_NORTH} & 10 & QSO cuts tuned for Bok/Mosaic data \\
\texttt{ELG\_LOP\_NORTH} & 11 & ELG at standard (ELG) priority tuned for Bok/Mosaic data \\
\texttt{ELG\_VLO\_NORTH} & 12 & Very-low priority ELG (filler) tuned for Bok/Mosaic data \\
\texttt{LRG\_SOUTH} & 16 & LRG cuts tuned for DECam data \\
\texttt{ELG\_SOUTH} & 17 & ELG cuts tuned for DECam data \\
\texttt{QSO\_SOUTH} & 18 & QSO cuts tuned for DECam data \\
\texttt{ELG\_LOP\_SOUTH} & 19 & ELG at standard (ELG) priority tuned for DECam data \\
\texttt{ELG\_VLO\_SOUTH} & 20 & Very-low priority ELG (filler) tuned for DECam data \\
\texttt{SKY} & 32 & Blank sky locations \\
\texttt{STD\_FAINT} & 33 & Standard stars for dark conditions \\
\texttt{STD\_WD} & 34 & White dwarf stars \\
\texttt{STD\_BRIGHT} & 35 & Standard stars for bright conditions \\
\texttt{BAD\_SKY} & 36 & Blank sky locations that are imperfect but still useable \\
\texttt{SUPP\_SKY} & 37 & SKY is based on Gaia-avoidance (SKY will be set, too) \\
\texttt{NO\_TARGET} & 49 & No known target at this location \\
\texttt{BRIGHT\_OBJECT} & 50 & Known bright object to avoid \\
\texttt{IN\_BRIGHT\_OBJECT} & 51 & Too near a bright object; DO NOT OBSERVE \\
\texttt{NEAR\_BRIGHT\_OBJECT} & 52 & Near a bright object but okay to observe \\
\texttt{BGS\_ANY} & 60 & Any BGS bit is set \\
\texttt{MWS\_ANY} & 61 & Any MWS bit is set \\
\texttt{SCND\_ANY} & 62 & Any secondary bit is set \\
\enddata
\begin{center}
\tablecomments{Bits are stored in the \texttt{desi\_mask} and accessed via the \texttt{DESI\_TARGET} column (for more details see \citealt{TS.Pipeline.Myers.2023}).}
\end{center}
\end{deluxetable*}

\begin{deluxetable*}{ccc}[htb]
\tablecaption{Bright Galaxy Survey (BGS) targeting bits for the DESI Main Survey.\label{table:mainbgs}}
\tablewidth{0pt}
\tablehead{
\colhead{Bit-name} &
\colhead{Bit-value} &
\colhead{Description}
}
\startdata
\texttt{BGS\_FAINT} & 0 & BGS faint targets \\
\texttt{BGS\_BRIGHT} & 1 & BGS bright targets \\
\texttt{BGS\_WISE} & 2 & BGS \textit{WISE} targets \\
\texttt{BGS\_FAINT\_HIP} & 3 & BGS faint targets at bright priority \\
\texttt{BGS\_FAINT\_NORTH} & 8 & BGS faint cuts tuned for Bok/Mosaic \\
\texttt{BGS\_BRIGHT\_NORTH} & 9 & BGS bright cuts tuned for Bok/Mosaic \\
\texttt{BGS\_WISE\_NORTH} & 10 & BGS \textit{WISE} cuts tuned for Bok/Mosaic \\
\texttt{BGS\_FAINT\_SOUTH} & 16 & BGS faint cuts tuned for DECam \\
\texttt{BGS\_BRIGHT\_SOUTH} & 17 & BGS bright cuts tuned for DECam \\
\texttt{BGS\_WISE\_SOUTH} & 18 & BGS \textit{WISE} cuts tuned for DECam \\
\enddata
\begin{center}
\tablecomments{Bits are stored in the \texttt{bgs\_mask} and accessed via the \texttt{BGS\_TARGET} column (see \citealt{TS.Pipeline.Myers.2023} for more details).}
\end{center}
\end{deluxetable*}

\begin{deluxetable*}{ccc}[htb]
\tablecaption{Milky Way Survey (MWS) targeting bits for the DESI Main Survey.\label{table:mainmws}}
\tablewidth{0pt}
\tablehead{
\colhead{Bit-name} &
\colhead{Bit-value} &
\colhead{Description}
}
\startdata
\texttt{MWS\_BROAD} & 0 & MWS magnitude-limited bulk sample \\
\texttt{MWS\_WD} & 1 & MWS white dwarf \\
\texttt{MWS\_NEARBY} & 2 & MWS volume-complete $\sim$100\,pc sample \\
\texttt{MWS\_BROAD\_NORTH} & 4 & MWS cuts tuned for Bok/Mosaic \\
\texttt{MWS\_BROAD\_SOUTH} & 5 & MWS cuts tuned for DECam \\
\texttt{MWS\_BHB} & 6 & MWS Blue Horizontal Branch stars \\
\texttt{MWS\_MAIN\_BLUE} & 8 & MWS magnitude-limited blue sample \\
\texttt{MWS\_MAIN\_BLUE\_NORTH} & 9 & MWS magnitude-limited blue sample tuned for Bok/Mosaic \\
\texttt{MWS\_MAIN\_BLUE\_SOUTH} & 10 & MWS magnitude-limited blue sample tuned for DECam \\
\texttt{MWS\_MAIN\_RED} & 11 & MWS magnitude-limited red sample \\
\texttt{MWS\_MAIN\_RED\_NORTH} & 12 & MWS magnitude-limited red sample tuned for Bok/Mosaic \\
\texttt{MWS\_MAIN\_RED\_SOUTH} & 13 & MWS magnitude-limited red sample tuned for DECam \\
\texttt{MWS\_FAINT\_BLUE} & 14 & MWS faint extension, blue side \\
\texttt{MWS\_FAINT\_BLUE\_NORTH} & 15 & MWS faint extension, blue side (Bok/Mosaic) \\
\texttt{MWS\_FAINT\_BLUE\_SOUTH} & 16 & MWS faint extension, blue side (DECam) \\
\texttt{MWS\_FAINT\_RED} & 17 & MWS faint extension, red side \\
\texttt{MWS\_FAINT\_RED\_NORTH} & 18 & MWS faint extension, red side (Bok/Mosaic) \\
\texttt{MWS\_FAINT\_RED\_SOUTH} & 19 & MWS faint extension, red side (DECam) \\
\texttt{GAIA\_STD\_FAINT} & 33 & Gaia-based standard stars for dark conditions \\
\texttt{GAIA\_STD\_WD} & 34 & Gaia-based white dwarf stars \\
\texttt{GAIA\_STD\_BRIGHT} & 35 & Gaia-based standard stars for bright conditions \\
\texttt{BACKUP\_GIANT\_LOP} & 58 & Giant candidate backup targets \\
\texttt{BACKUP\_GIANT} & 59 & Giant candidate backup targets \\
\texttt{BACKUP\_BRIGHT} & 60 & Bright backup Gaia targets \\
\texttt{BACKUP\_FAINT} & 61 & Fainter backup Gaia targets \\
\texttt{BACKUP\_VERY\_FAINT} & 62 & Even fainter backup Gaia targets \\
\enddata
\begin{center}
\tablecomments{Bits are stored in the \texttt{mws\_mask} and accessed via the \texttt{MWS\_TARGET} column  (see \citealt{TS.Pipeline.Myers.2023} for more details).}
\end{center}
\end{deluxetable*}

\section{Secondary Targets}\label{app:secondarytargets}


In addition to its primary science goals, the DESI survey incorporates a range of ``secondary'' targets to pursue bespoke research (see \S\ref{sec:targets}). In this appendix, we describe the secondary target campaigns included in the main survey and outline how the bit-values in their \texttt{scnd\_mask} and \texttt{SCND\_TARGET} column \citep[see \S2.4 of][]{TS.Pipeline.Myers.2023} can be linked back to the relevant program.\footnote{Bitmasks such as \texttt{SV3\_SCND\_TARGET} \citep{TS.Pipeline.Myers.2023} are also typically present in the fibermap, but will be zero for main-survey targets. Similarly, \texttt{SCND\_TARGET} will be zero for other survey types.}

In Table~\ref{table:mainsec} we list the bit-names and bit-values for secondary targets that were scheduled for observation during the DESI main survey. The vast majority of these secondary targeting programs are documented in Appendix~B of \citet{DESI2023b.KP1.EDR}. Below, we describe the \textit{new} programs that were added for the main survey. Further scientific justification for many of these new programs can be found in the MWS overview paper \citep{MWS.TS.Cooper.2023}. Details about the selection of each type of secondary target are also available at the associated \texttt{docs} link for main survey secondary targets.\footnote{\url{https://data.desi.lbl.gov/public/edr/target/secondary/main/docs}}

\begin{deluxetable*}{cccc}[t]
\tablecaption{Secondary targeting bits for the DESI Main Survey.\label{table:mainsec}}
\tablewidth{0pt}
\tablehead{
\colhead{Bit-name} & 
\colhead{Bit-value} & 
\colhead{Bit-name} & 
\colhead{Bit-value}
}
\startdata
\texttt{VETO} & 0 & \texttt{WISE\_VAR\_QSO} & 35 \\
\texttt{UDG} & 1 & \texttt{Z5\_QSO} & 36 \\
\texttt{FIRST\_MALS} & 2 & \texttt{MWS\_RR\_LYRAE} & 37 \\
\texttt{QSO\_RED} & 5 & \texttt{MWS\_MAIN\_CLUSTER\_SV} & 38 \\
\texttt{MWS\_CLUS\_GAL\_DEEP} & 10 & \texttt{BRIGHT\_HPM} & 40 \\
\texttt{LOW\_MASS\_AGN} & 11 & \texttt{WD\_BINARIES\_BRIGHT} & 41 \\
\texttt{FAINT\_HPM} & 12 & \texttt{WD\_BINARIES\_DARK} & 42 \\
\texttt{LOW\_Z\_TIER1} & 15 & \texttt{PV\_BRIGHT\_HIGH} & 43 \\
\texttt{LOW\_Z\_TIER2} & 16 & \texttt{PV\_BRIGHT\_MEDIUM} & 44 \\
\texttt{LOW\_Z\_TIER3} & 17 & \texttt{PV\_BRIGHT\_LOW} & 45 \\
\texttt{BHB} & 18 & \texttt{PV\_DARK\_HIGH} & 46 \\
\texttt{SPCV} & 19 & \texttt{PV\_DARK\_MEDIUM} & 47 \\
\texttt{DC3R2\_GAMA} & 20 & \texttt{PV\_DARK\_LOW} & 48 \\
\texttt{PSF\_OUT\_BRIGHT} & 25 & \texttt{GC\_BRIGHT} & 49 \\
\texttt{PSF\_OUT\_DARK} & 26 & \texttt{GC\_DARK} & 50 \\
\texttt{HPM\_SOUM} & 27 & \texttt{DWF\_BRIGHT\_HI} & 51 \\
\texttt{SN\_HOSTS} & 28 & \texttt{DWF\_BRIGHT\_LO} & 52 \\
\texttt{GAL\_CLUS\_BCG} & 29 & \texttt{DWF\_DARK\_HI} & 53 \\
\texttt{GAL\_CLUS\_2ND} & 30 & \texttt{DWF\_DARK\_LO} & 54 \\
\texttt{GAL\_CLUS\_SAT} & 31 & \texttt{BRIGHT\_TOO\_LOP} & 59 \\
\texttt{MWS\_FAINT\_BLUE} & 32 & \texttt{BRIGHT\_TOO\_HIP} & 60 \\
\texttt{MWS\_FAINT\_RED} & 33 & \texttt{DARK\_TOO\_LOP} & 61 \\
\texttt{STRONG\_LENS} & 34 & \texttt{DARK\_TOO\_HIP} & 62 \\
\enddata
\begin{center}
\tablecomments{Bits are stored in the \texttt{scnd\_mask} and accessed via the \texttt{SCND\_TARGET} column (see \citealt{TS.Pipeline.Myers.2023} for more details).}
\end{center}
\end{deluxetable*}

\subsection{\texttt{MWS\_FAINT\_BLUE}, \texttt{MWS\_FAINT\_RED}}

The \texttt{MWS_FAINT} bits denote former primary target classes that were eventually implemented as secondary target classes due to the bug described in \S5.1 of \citet{TS.Pipeline.Myers.2023}. These target classes are detailed extensively in \citet{MWS.TS.Cooper.2023}.

\subsection{\texttt{MWS\_RR\_LYRAE}}

This bit denotes a target class that is identical to the Gaia DR2 RR Lyrae variable targets (\texttt{MWS\_RR\_LYR}) described in Appendix B of \citet{DESI2023b.KP1.EDR}. These RR Lyrae targets were observed during Target Selection Validation (\texttt{SURVEY=sv1}) but needed to be scheduled at a higher priority for the main survey. Because priority is a \textit{bit}-specific feature in the context of the \texttt{desitarget} code, a new bit was required to denote the same target class.

\subsection{\texttt{GC\_BRIGHT}, \texttt{GC\_DARK}}
\label{sec:GC}

Primary target selections typically mask out sources near globular clusters using the Legacy Surveys \texttt{CLUSTER} bit-mask.\footnote{https://www.legacysurvey.org/dr9/bitmasks/} The \texttt{GC} secondary programs are designed to select stars in globular clusters to ameliorate any paucity of DESI targets in these \texttt{CLUSTER} regions. The targets consist of 
high-probability ($P > 0.3$) cluster members from \citet{vasiliev21a}. As the \texttt{GC} targets have a very low sky density ($< 5\, \rm{deg}^2$) they are unlikely to interfere with DESI large scale structure analyses, so they are prioritized just below DESI primary targets and white dwarf targets. The \texttt{GC\_BRIGHT} (\texttt{GC\_DARK}) sample, intended for observations during bright (dark) time, is limited to $16 < r < 20$ ($19 < r < 21)$.

\subsection{\texttt{DWF\_BRIGHT\_HI}, \texttt{DWF\_BRIGHT\_LO}, \texttt{DWF\_DARK\_HI}, \texttt{DWF\_DARK\_LO}}

The \texttt{DWF} programs are philosophically similar to the \texttt{GC} programs outlined in \S\ref{sec:GC}, but are designed to target Milky Way satellite galaxies rather than globular clusters. As is the case for the \texttt{GC} programs, the \texttt{DWF} targets cover a relatively small sky area ($\sim 12\, \rm{deg}^2$), so are prioritized just below primary targets. The \texttt{DWF\_BRIGHT\_HI} and \texttt{DWF\_DARK\_HI} target classes are scheduled at slightly higher priority than the \texttt{DWF\_BRIGHT\_LO} and \texttt{DWF\_DARK\_LO} target classes, which, in turn share the priority of the \texttt{GC} programs. The target sample comprises members of known Milky Way satellite galaxies. The galaxies (see Table~\ref{table:dwf}) are sub-selected to have a $V$-band absolute magnitude $> -10$~mag, a declination north of 30$^\circ$, a half-light-radius of $< 30\arcmin$, at least 5 member stars in Gaia, and coverage in LS/DR9.

A star is considered a ``member'' of the relevant galaxy if it is 
within five (three) half-light-radii of the galactic center 
if the galaxy's half-light-radius is $< (>)~7\arcmin$. The selection is further refined by cross-matching to Gaia EDR3 \citep{gaia-collaboration22a} and restricting to likely members based on parallax and proper motion. For stars that are not in Gaia EDR3, imaging from LS/DR9 is used to estimate whether a target is a member of the appropriate galaxy. A star is assigned to be ``high'' (\texttt{HI}) or ``low'' (\texttt{LO}) priority based on whether information from Gaia and the Legacy Surveys implies that it is highly likely to be a member of the appropriate dwarf galaxy. We list the resulting numbers of targets and the area covered by each galaxy in Table~\ref{table:dwf}. Finally, targets are assigned to the \texttt{BRIGHT} or \texttt{DARK} class, intended for observations during bright or dark time if they fall in the magnitude range $16 < r < 20$ or $19 < r < 21$, respectively.

\begin{deluxetable*}{cccc}[t]
\tablecaption{Milky Way satellites in the \texttt{DWF} program.\label{table:dwf}}
\tablewidth{0pt}
\tablehead{
\colhead{Dwarf} & 
\colhead{} & 
\colhead{} & 
\colhead{Area} \\
Galaxy & $N_{\rm high}$ & $N_{\rm low}$ & \colhead{(${\rm deg}^2$)}}
\startdata
          aquarius\_2   &      16    &     268  &  0.57    \\
            bootes\_1   &     171    &     358  &  0.78    \\
            bootes\_2   &      22    &      56  &  0.22    \\
    canes\_venatici\_1   &     141    &     231  &  0.40    \\
    canes\_venatici\_2   &      15    &      24  &  0.05    \\
           columba\_1   &       6    &      40  &  0.11    \\
    coma\_berenices\_1   &      41    &     176  &  0.69    \\
             draco\_1   &    1653    &    1913  &  0.73    \\
             draco\_2   &      25    &      59  &  0.20    \\
          hercules\_1   &      44    &     434  &  0.69    \\
               leo\_2   &     342    &     531  &  0.14    \\
               leo\_4   &       8    &      63  &  0.14    \\
               leo\_5   &       7    &      11  &  0.02    \\
             segue\_1   &      25    &     122  &  0.29    \\
             segue\_2   &      19    &     117  &  0.31    \\
           sextans\_1   &    1450    &    2482  &  2.14    \\
        ursa\_major\_1   &      52    &     116  &  0.54    \\
        ursa\_major\_2   &      52    &     323  &  1.50    \\
        ursa\_minor\_1   &    2032    &    2680  &  2.63    \\
           willman\_1   &       8    &      27  &  0.14    \\
\enddata
\begin{center}
\tablecomments{$N_{\rm high}$ and $N_{\rm low}$ refer to the number of targets in the high-priority (\texttt{DWF\_BRIGHT\_HI}, \texttt{DWF\_DARK\_HI}) and low-priority (\texttt{DWF\_BRIGHT\_LO}, \texttt{DWF\_BRIGHT\_LO}) programs, respectively.}
\end{center}
\end{deluxetable*}

\section{Special Observations}\label{app:tertiarytargets}

As discussed in \S\ref{sec:special}, DR1 contains a number of tiles that were dedicated to special observations. These tiles can be identified by having \texttt{special} in the \texttt{SURVEY} column of the DR1 \texttt{tiles-iron.fits} file (see \S\ref{sec:products}). The vast majority of science targets for these programs have the \texttt{SCND\_ANY} bit set in \texttt{desi\_mask} (see Table~\ref{table:maindark}) and the \texttt{DARK_TOO_LOP} or \texttt{BRIGHT\_TOO\_LOP} bit set in \texttt{scnd\_mask} (see Table~\ref{table:mainsec}). This is because science targets for rapid-turnaround special tiles are all loosely interpreted to be ``Targets of Opportunity.'' In this appendix, we briefly describe the DR1 special observations summarized in Table~\ref{tab:specialobs}.



\subsection{\texttt{backup}}

The special \texttt{backup} tiles were designed to test the backup program target selection algorithms and implementation before enabling routine observations of the backup program in the main survey \citep{MWS.TS.Cooper.2023, dey25a}). These tiles may use different versions of targeting than the eventual main survey backup tiles; consequently, we do not recommend combining special and main survey backup tiles.

\subsection{\texttt{bright}}

Multiple test tiles were taken in the context of the \texttt{bright} special program for different purposes: 

\begin{itemize}

\item Tiles 80978 and 80979 were designed to test how successfully DESI could acquire redshifts in very poor conditions, and duplicated the targets observed on standard bright tiles 20655 and 21071. The standard bright tiles and special-program bright tiles were assigned different \texttt{TILEID}s to prevent inadvertently combining normal good data with the special (intentionally poor) data acquired as part of the special program tiles. 

\item Tiles 80980 and 80981 were designed to include Targets of Opportunity \citep[see, e.g. \S3.2.2 of][]{TS.Pipeline.Myers.2023} in addition to normal bright targets, in order to test the infrastructure for handling candidate transient sources. 

\item Tiles 82258--82268 were designed with a variety of updates to the DESI fiber assignment code to attempt to optimize the density of DESI fibers assigned to targets.

\end{itemize}

\subsection{\texttt{dark}}

The special \texttt{dark} program also covered tiles with a range of different purposes:

\begin{itemize}
    \item Tile 80977 was a deep tile designed to study the nature of QSO and ELG contaminants in the Sagittarius Stream \citep[e.g.][]{lynden-bell95a}. 
    
    \item Tiles 81100 and 81112 were test tiles used to validate the (ultimately adopted) process by which DESI uses nonfunctional positioners to help measure the sky background \citep[see, e.g., \S 5.6 of][]{SurveyOps.Schlafly.2023}.
    
    \item Tile 82237 was observed after DESI was shut down in the summer of 2021 to test preparedness for returning to normal operations.

\end{itemize}

\subsection{\texttt{m31}}

The \texttt{m31} special tiles were designed to study the properties of the halo of M31. These observations are detailed in \citet{dey23a}.

\subsection{\texttt{odin}}

The \texttt{odin} special tile targeted Ly$\alpha$ emitters (LAEs) and Lyman-break galaxies (LBGs) selected in the COSMOS field \citep{scoville07a}. The scientific aim of this program was to investigate the capability of DESI to execute future, higher redshift surveys. 


\subsection{\texttt{tertiary1}}

The \texttt{tertiary1} special tiles obtained spectroscopy for a large fraction of magnitude $z < 21.6$ galaxies in the COSMOS field, in order to provide a very dense and complete set of galaxy redshifts. This program marked the transition from using a variety of names to describe special tiles to incrementing each new special program as a ``tertiary" program \citep[see, e.g.,][]{TS.Pipeline.Myers.2023}.

\section{Known Issues}\label{app:issues}

After the Iron production was finished, several problems were found in the coaddition of target metadata, impacting the \texttt{FIBERMAP} tables in \texttt{coadd*.fits} and \texttt{redrock*.fits} files, which are further propagated into the summary redshift catalogs in \texttt{spectro/redux/iron/zcatalog/v0/}. Note that the underlying spectra, coadded spectra, and redshift fits are not impacted by the following issues, only the metadata about those spectra are impacted. These issues have been fixed in all summary catalogs in \texttt{spectro/redux/iron/zcatalog/v1/}, including both tile- and HEALPix-based versions, and will be correct in future data releases. They have \textit{not} been fixed in the $\sim$200k individual \texttt{coadd} and \texttt{redrock} files. 

The following issues in the \texttt{FIBERMAP} metadata are resolved in the \texttt{v1} catalogs, but not in \texttt{v0} or the individual files:

\begin{itemize}

\item 0.03\% of targets incorrectly have \texttt{COADD\_FIBERSTATUS=0}
even though all of their data are masked.  These targets will correctly have \texttt{ZWARN} bit 9 (\texttt{NODATA}) set in the redshift fits. Quality cuts based solely upon \texttt{COADD\_FIBERSTATUS} without also cutting on \texttt{ZWARN} will have some contamination. This amounts to roughly 8,500 HEALPix-based entries in the individual files. Users who need \texttt{COADD\_FIBERSTATUS} should get that information from the appropriate \texttt{v1} summary catalog by matching on \texttt{TARGETID}.

\item \texttt{MEAN\_FIBER\_\{RA,DEC\}}, \texttt{STD\_FIBER\_\{RA,DEC\}}, \texttt{MEAN\_DELTA\_\{X,Y\}}, \texttt{RMS\_DELTA\_\{X,Y\}}, and \texttt{MEAN\_PSF\_TO\_FIBER\_SPECFLUX} incorrectly included
all exposures, not just those that passed the quality cuts and were used in the
coadd.  This issue impacts approximately 0.7\% of targets. This amounts to roughly 200,000 HEALPix-based entries in the individual files. Users who need these quantities should get that information from the appropriate \texttt{v1} summary catalog by matching on \texttt{TARGETID}.

\item \texttt{STD\_FIBER\_RA}, the standard deviation of contributing \texttt{FIBER_RA} values, is incorrect for $\sim$17\% of the targets. This amounts to roughly 4.75 million HEALPix-based entries in the individual files. It only impacts objects that were observed on more than one night, and all \texttt{FIBER\_RA} values are correct. Users requiring \texttt{STD\_FIBER\_RA} should get that information from the appropriate \texttt{v1} summary catalog by matching on \texttt{TARGETID}. 

\item Targets that were originally observed under one survey/program and then were later
reselected as secondary targets are missing their secondary target bits in the coadded
\texttt{DESI\_TARGET}, \texttt{SCND\_TARGET}, \texttt{SV1\_DESI\_TARGET}, \texttt{SV1\_MWS\_TARGET}, and \texttt{SV1\_SCND\_TARGET} bit masks.  This issue impacts $\sim$0.0006\% of targets. This amounts to less than 200 HEALPix-based entries in the individual files. Those requiring these quantities should get that information from the appropriate \texttt{v1} summary catalog by matching on \texttt{TARGETID}.

\end{itemize}

Other known issues that have \textit{not} been resolved for DR1 include:
\begin{itemize}

\item For both EDR and DR1 the coadd resolution matrices are suboptimal due to mis-weighting of the individual resolution matrices. However, these effects only affect spectra that were coadded from more than one exposure and are small, with $\sim1.7$\% of pixels having resolution matrix elements incorrect by more than 0.01, and $\sim0.05$\% of pixels incorrect by more than 0.1. This is generally insignificant but can impact detailed studies of spectra. This is now fixed in the latest version of \texttt{desispec}, which can be used with the Iron \texttt{spectra} files to generate new \texttt{coadd} files with the correct resolution matrices.

\item There was a bug in inverse variance estimation during flux calibration of all spectra, which caused the variance of all spectra to be underestimated. The effects are negligible at low signal-to-noise ratio (S/N) per pixel, but become significant at around S/N $\sim20-30$. At a S/N per pixel of $\sim100$, the underestimation of the uncertainties can be as much as a factor of 5.

\item Radial velocity systematics in DR1 main survey backup program data have been identified with a scatter of 2-3 km/s, which occasionally reach values up to 20 km/s, due to limitations in the wavelength calibration during Guadalupe and Iron processing.

\item We did not properly account for a change in the string representation of Boolean values between Gaia DR2 and EDR3 \citep{lindegren18a,gaia-collaboration22a}. Due to this oversight, the Gaia \texttt{DUPLICATED_SOURCE} column is always \texttt{False} for DESI targets that derived their values from EDR3. The implication of this is that no source in DESI targeted using Gaia EDR3 (see \S4.1.4 of \citealt{TS.Pipeline.Myers.2023}) would have been masked on \texttt{DUPLICATED_SOURCE}. Almost 2 million sources in the Iron HEALPix-based catalog were selected using Gaia EDR3. Approximately 2.8\% of sources in Gaia EDR3 have the {\tt DUPLICATED_SOURCE} column set, so roughly $\approx$56k of the $\approx$2M dark-time sources in DR1 may have been unintentionally targeted.

\item The QuasarNet afterburner provides \texttt{Z_NEW} computed from Redrock with only the quasar templates using a uniform prior of $\pm 0.05$ about its coarse estimated \texttt{Z_QN}. When the redshift prior overlaps with $z>1.4$, only the high redshift quasar templates were used. This causes a slight excess of quasars with \texttt{Z_NEW}~$\approx1.4$ (the high redshift template lower bound) and an increased \texttt{ZWARN} rate for quasars with $1.2 < z < 1.8$. This is propagated to the \texttt{Z} column in the LSS catalogs, impacting $\sim0.5$\% of the quasars in the corresponding redshift range.

\item Redshifts for Ly$\alpha$ quasars are systematically underestimated with increasing redshift owing to incorrect modeling of the Lyman series optical depth in Redrock, also present for EDR \citep{wu23a, KP6s4-Bault}. The average bias is modest at $z \approx 2$, but exceeds 100 km/s at $z>3$. This will be corrected in future releases, but remains present in Iron. The DR1 ZLyA VAC provides corrected redshifts for Ly$\alpha$ quasars.

\item A calibration problem on petal/spectrograph 9 on night 20211212
led to many incorrect redshifts for fibers 4500--4999 on the 17 tiles observed on that night (8500 spectra, $\sim$0.03\% of Iron).
DR1 includes a supplementary release \texttt{spectro/redux/reproc_20211212_iron/}, produced using the same code tags as Iron but with a better choice of calibration data. All 17 tiles were reprocessed and provided as tile-based redshifts. This includes tile 7733 which was re-observed on 20211215 and thus included in the cumulative coadds for 20211215. All other impacted tiles were last observed on 20211212.  HEALPix-based coadds and summary catalogs were not produced for this mini production. The reprocessed data were used for the LSS catalogs, but were not used to replace values in Iron itself.

\item \texttt{FIBER\_\{RA,DEC\}} = \texttt{FIBER\_\{X,Y\}} = 0.0 for broken fibers (whose locations cannot be measured)
and for fibers whose positions were not measured by the Fiber View Camera on individual exposures.
These cases have \texttt{FIBERSTATUS} bit 8 \texttt{MISSINGPOSITION} set.

\item In early main survey data, some non-target \texttt{BAD\_SKY} locations near bright objects received the same \texttt{TARGETID} despite having different (RA,Dec) locations, and vice-a-versa (i.e., different \texttt{TARGETID} for the same (RA,Dec) on different overlapping tiles).
This issue only impacts non-science, non-calibration spectra when a positioner was unable to reach a good target.

\item \texttt{DESINAME}, which gives an identifier for DESI objects based on sky location truncated to a precision of $10^{-4}$ arcsseconds, is incorrect in the catalogs and individual files for $-0.1 < \texttt{TARGET_DEC} < -0.0001$. Correct \texttt{DESINAME} values can be generated using the modern version of \texttt{desiutil} by calling the function \texttt{desiutil.names.radec_to_desiname} with the object's \texttt{TARGET_RA} and \texttt{TARGET_DEC}. This issue led to 121,132 incorrect names in the HEALPix-based catalog; or 0.4\% of Iron.

\item Some stuck positioners from Survey Validation data with \texttt{TARGETID<0} appear in the tile-based redshift catalogs, but not in the HEALPix-based redshift catalogs.

\item Tile 21917 (main survey bright program) has valid \texttt{FIBER\_\{X,Y\}} values, but is missing the focal plane $(x,y) \rightarrow (\textrm{RA},\textrm{Dec})$ transform in the raw data and thus has \texttt{FIBER\_\{RA,DEC\}} = 0.0.
In this case, the \texttt{TARGET\_RA} and \texttt{TARGET\_DEC} values record the intended target position and the \texttt{FIBER\_\{X,Y\}} and \texttt{DELTA\_\{X,Y\}} values remain valid and were used for quality cuts to set \texttt{FIBERSTATUS} bits 9 \texttt{BADPOSITION} and 10 \texttt{POORPOSITION} if the fiber was away from the intended target.

\end{itemize}

Finally, the following known issues were present in the previous data release (EDR) and documented in \citet{DESI2023b.KP1.EDR}. They remain issues in DR1 and are included here for completeness:

\begin{itemize}

\item Redrock templates do not include Active Galactic Nucleus (AGN)-like galaxies with a mixture of broad and narrow lines.  As a result, these types of galaxies are often fit equally well (or equally poorly) with either \texttt{GALAXY} or \texttt{QSO} templates at the same redshift, which can also trigger \texttt{ZWARN} bit 2 (value $2^2=4$) for \texttt{LOW_DELTACHI2} since the $\chi^2$ difference between the two fits is small, indicating an ambiguous answer.

\item There are cases where Redrock is overconfident and reports \texttt{ZWARN=0}, i.e., no known problems, even though the fit is incorrect.  This can include unphysical fits due to the over-flexibility of PCA template linear combinations. This is particularly true for sky fibers which have a higher fraction of \texttt{ZWARN=0} than would be expected from purely random fluctuations.  Users should be especially cautious in any search for serendipitous targets in nominally blank sky fibers.

\item The Redrock galaxy fits extend to redshift $z=1.7$, though the range $1.6 < z < 1.63$ is only constrained by the [\ion{O}{2}] doublet in the midst of significant sky background and $1.63 < z < 1.7$ has no major emission line coverage.
Thus $1.6 < z < 1.7$ is particularly susceptible to unphysical fits.  This was the motivation for the LSS catalogs to only consider galaxies with $z<1.6$.

\item For most of the tiles in Target Selection Validation (SV1), proper-motion corrections were applied in Fiberassign when the tile was designed.\footnote{The design date can differ from when a tile was observed.} A consequence is that the (\texttt{TARGET\_RA}, \texttt{TARGET\_DEC}, and \texttt{REF\_EPOCH}) values are altered to have a \texttt{REF\_EPOCH} of the date that the tile was designed, which makes them differ from the input photometric column values. The information is correct and consistent with the photometry, however.

\end{itemize}

Additional known issues and clarifications will be documented at \url{https://data.desi.lbl.gov/doc/releases/dr1} as they arise.

\section{Summaries of Value Added Catalogs}\label{app:vacs}

This appendix expands upon the set of VACs introduced in \S\ref{sec:vac-files} and summarized in Table~\ref{tab:vacs}. Specifically, Appendix~\ref{sec:vac_general} contains general VACs not tied to a specific science case and spanning a variety of scientific applications; Appendix~\ref{sec:vac_mws} contains VACs that are primarily concerned with Milky Way or stellar science; Appendix~\ref{sec:vac_extragalactic} contains VACs associated with extragalactic science, including voids, galaxy groups, galaxy statistics and strong lenses; Appendix~\ref{sec:vac_quasar} contains VACs associated with quasar and active galactic nuclei (AGN) science, including summary statistics, updated redshifts, \ion{Mg}{2} absorption systems and damped Ly$\alpha$ (DLA) systems; and Appendix~\ref{sec:vac_lya} contains data products produced by the Ly$\alpha$ Forest Working Group which were used in the analysis in \cite{DESI2024.IV.KP6}.

Some of the VACs described below are finalized and will not change (e.g., those with published companion papers), while others associated with papers currently in preparation may be updated after the public data release. The most current version of each VAC and its associated documentation can always be found at the DESI VAC portal.\footnote{\url{https://data.desi.lbl.gov/doc/vac}}

\subsection{General VACs}\label{sec:vac_general}

\underline{LS/DR9 Photometry VAC} delivers merged targeting catalogs (\textit{targetphot}) and Tractor\footnote{\url{https://github.com/dstndstn/tractor}} \citep{lang16a} catalog photometry (\textit{tractorphot}) from the  Legacy Surveys DR9 \citep[LS/DR9\footnote{\url{https://www.legacysurvey.org/dr9}};][]{LS.Overview.Dey.2019} for all observed and potential targets (excluding sky fibers) in DR1.\footnote{\url{https://github.com/moustakas/desi-photometry}} The observed targets in this VAC correspond to objects with at least one observation in DR1, while the potential targets are the targets that DESI \textit{could have} observed in a given fiber-assignment configuration (including the objects that were actually observed).

\underline{Sky Spectra VAC} provides over 9,000 sky spectra from DR1, accompanied by detailed metadata, derived from the DESI spectroscopic pipeline. These spectra are essential for understanding the sky background, enabling the identification of natural and artificial emission features, and improving the analysis of astronomical targets. The metadata include comprehensive observational parameters, such as exposure time, airmass, Galactic extinction, moon and sun positions, and atmospheric conditions, offering a rich context for each spectrum. The primary purpose of this catalog is to be a resource for studies of atmospheric phenomena, light pollution, and calibration.

\underline{BAO Cosmology Results VAC} provides the cosmology chains and posterior maximization results derived from the BAO analysis in \cite{DESI2024.VI.KP7A}.

\underline{Full Shape Cosmology Results VAC} provides the cosmology chains and posterior maximization results used full shape analysis in \cite{DESI2024.VII.KP7B}.

\subsection{Milky Way Survey VACs}\label{sec:vac_mws}

\underline{MWS VAC} provides analysis of stellar spectra by the Milky Way Survey Working Group \citep{koposov25a}. The analysis involves fitting DESI spectra using two pipelines: RVSpecFit \citep{koposov24a} and FERRE \citep{allende-prieto23a} (see \citealt{MWS.TS.Cooper.2023} and \citealt{koposov24a} for more details). The RVSpecFit pipeline derives radial velocity measurements, stellar parameters, and abundances, while FERRE provides stellar parameters and abundance measurements. Both pipelines have been applied to approximately 6.5M coadded spectra of stars or possible stars from DESI DR1. Additionally, the RVSpecFit pipeline was run on approximately 9.5M single-exposure spectra, enabling radial velocity variability analyses. Since the DESI EDR, both pipelines have been significantly improved, with RVSpecFit now using neural networks to interpolate stellar spectra, thus eliminating issues with clustering of stellar parameters on grid nodes that was present in the EDR VAC.
The data products in this VAC include tables of stellar parameters, radial velocities, their uncertainties, and the best-fit models for the DESI spectra. In addition, the data tables have been cross-matched with the latest Gaia DR3 release.

\underline{MWS BHB VAC} contains spectroscopically confirmed BHB stars. BHB stars are excellent tracers of the Milky Way halo; they are intrinsically bright, so they can be observed to large distances, and there exists a simple polynomial relationship between their color and absolute magnitude \citep{deason11a, belokurov16a}, meaning that their distances can be computed with high precision. BHB candidates were targeted in the MWS \citep[see \S4.4.4 in ][]{MWS.TS.Cooper.2023} and as secondary targets observed in dark time \citep[see][]{TS.Pipeline.Myers.2023}. Due to photometric uncertainties, targeted BHB candidates contain contamination primarily from more nearby blue straggler stars. The DESI spectroscopic observations, however, allow us to remove this contamination. The BHB VAC contains over 6,300 spectroscopically confirmed BHB stars, all having distances derived from photometry, as described in \citet{bystrom25a}. This BHB catalogue contains the stellar parameters from the RVSpecFit pipeline \citep{koposov24a}, including radial velocity, surface gravity and effective temperature, and reaches heliocentric distances of about 120~kpc.

\underline{MWS SpecDis VAC} provides distance measurements for over 4M stars from DR1 following the selection criteria outlined in \citet{li25a}. The VAC is based on a  data-driven approach which adopts a neural network to establish the connection between stellar spectra and luminosity. It is trained on Gaia parallax and Gaia $G$-band apparent magnitude measurements, with the following key improvements: (1) We adopt a special training label to avoid cutting off negative parallaxes; (2) We do not include any parallax uncertainty cuts for the training sample, but we include the parallax uncertainties in the loss function. This approach ensures that the training sample does not suffer from biases in parallax, and our training sample includes a significant number of distant giants. We can achieve significant improvements in our measurements than Gaia parallaxes beyond 7~kpc. (3) We adopt principal component analysis to decrease the noise and the dimensionality of the spectra. In general, the distance uncertainties decrease with increasing average signal-to-noise ratio in the B and R arms of the DESI stellar spectra (see Table~\ref{tab:instrument_parameters}). We adopt a Gaussian mixture model to identify candidate binary stars. The final catalog offers distance, distance uncertainty, the binary candidate flag plus other useful photometric and astrometric information, cross-matched from the RVSpecFit values reported in the MWS VAC \citep{koposov24a} and Gaia DR3.

\underline{SPDist VAC} provides spectrophotometric distances in the form of absolute magnitude in the Gaia $G$-band for all stars observed by the DESI MWS, using a data-driven approach similar to \citet{thomas22a}. Distances are estimated using a fully connected multi-layer perceptron (MLP) model using Gaia DR3 photometry \citep{gaia-collaboration23a}, spectroscopic metallicity from the SP pipeline, and atmospheric parameters (effective temperature and surface gravity) derived from a combination of values from both the RV and SP pipelines \citep{koposov24a} as inputs. The MLP outputs the absolute magnitude in the Gaia $G$-band, which is subsequently used to calculate distances. Some stars exhibit significant discrepancies in spectroscopic parameters derived by the RV and SP pipelines, likely leading to unreliable distance estimates. These suspicious stars are flagged with the \texttt{FLAG\_GOOD} parameter. We recommend using this flag along with the quality flags recommended for the RV and SP pipelines \citep{koposov24a}. For distance modulus computations, we recommend using the median of the absolute magnitude distribution (\texttt{MG\_50}), while uncertainties can be derived from the 16th and 84th percentiles of the distribution. These uncertainties should be combined in quadrature with the intrinsic precision of the method (0.167 mag, 8\% relative distance precision) to obtain the total uncertainty. The primary difference between the methodologies of SpecDist and SPdist is that SpecDist predicts distances from the stellar spectra, whereas SPdist predicts distances from a full list of stellar parameters by the MWS RV and SP pipelines. A forthcoming paper will present a detailed comparison of the two methods.

\underline{Stellar Reddening VAC} contains the observed spectra and the RVSpecFit model spectra of stars that provided reddening measurements for the DESI dust map \citep{zhou24a}. The DESI dust map, publicly released with \citet{zhou24a}, is used in the DESI DR1 cosmology analysis \citep{DESI2024.II.KP3}. Also included in this VAC is the stellar catalog containing the per-star reddening measurements in the Dark Energy Camera \citep[DECam;][]{flaugher15a} filters as well as other properties from the DESI pipeline and the imaging catalogs. While the paper uses the first two years of DESI data, only the DR1 portion of the data is released in the VAC. Also note that while both the MWS VAC and this VAC use RVSpecFit and are based on the same DESI data, there are some key differences: (1) the RVSpecFit model spectra in this VAC are the zero-extinction spectra, whereas the MWS VAC model spectra include an additional multiplicative term to fit the smooth-varying component of the observed spectra; and (2) the DR1 MWS VAC uses a different (newer) version of RVSpecFit than the version used here. For general purposes (other than for studying dust), the MWS VAC should be used instead of this VAC.

\subsection{Extragalactic Science VACs}\label{sec:vac_extragalactic}


\underline{HETDEX VAC} contains HETDEX \citep{hill21a, gebhardt21a} and DESI spectra of HETDEX-selected Ly$\alpha$ emitter candidates \citep{mentuch-cooper23a} followed up by DESI. The VAC also contains the emission line fits from both sets of spectra. Details of the analysis can be found in \citet{landriau25a}.

\underline{DESIVAST VAC} contains cosmic voids identified within the DESI DR1 volume. The void locations are computed using a volume-limited subsample of the BGS Bright sample (see \citealt{BGS.TS.Hahn.2023}) extending to $z<0.24$. We apply evolutionary corrections to 479,486 galaxies and enforce a magnitude cut of $M_r < -20$. We then apply three void-finding algorithms: VoidFinder \citep{el-ad97a, hoyle02a, douglass22a}, V$^2$/VIDE \citep{neyrinck08a, sutter15a}, and V$^2$/REVOLVER \citep{nadathur19a}, to obtain three comparable void catalogs. Our VoidFinder catalog contains 3,765 voids, with 1,489 non-edge voids defined as those not bordering on the survey edges. For V$^2$/VIDE (V$^2$/REVOLVER), we find 1,478 (1,992) voids with 297 (389) non-edge voids. Further information regarding this VAC can be found in \citet{rincon25a}.

\underline{Dwarf Galaxy VAC} presents a sample of extragalactic dwarf galaxies ($M_{\text{star}} < 10^9 M_{\text{sun}}$) identified in DESI DR1. The catalog includes galaxies selected from the BGS and ELG samples, as well as the LOW-Z secondary target program \citep{darragh-ford23a}, spanning a redshift range of $0.001 < z < 0.5$. However, due to differences in target selection criteria across these samples, the catalog is not uniformly selected. Stellar masses are derived using CIGALE\footnote{\url{https://cigale.lam.fr}} \citep{boquien19a, siudek24a} and optical color-based prescriptions \citep{mao24a, de-los-reyes24a}. In a forthcoming paper (Manwadkar et al. 2026, in preparation), we will supplement this catalog with information on spectroscopic and photometric completeness, and an analysis of quenched fractions as a function of environment.

\underline{EmFit VAC} provides emission-line fitting results using the EmFit code \citep{pucha25a} run on low-redshift ($z\leq0.45$) galaxy spectra, focusing on eight nebular emission lines and doublets: H$\beta$, [\ion{O}{3}]~$\lambda\lambda$4959,5007, [\ion{N}{2}]~$\lambda\lambda$6548,83, H$\alpha$, and [\ion{S}{2}]~$\lambda\lambda$6716,31. The code tests for the presence of extra components in the [\ion{S}{2}] and [\ion{O}{3}] doublets independently, and the profiles of the rest of the narrow components are designed to match the profile of the [\ion{S}{2}] lines. It further tests for the presence of a possible broad component in the Balmer lines. The flux and width measurements along with their uncertainties for all detected components are reported in this catalog \citep{pucha25a}.

\underline{FastSpecfit VAC} delivers a broad range of observed-frame, rest-frame, and intrinsic physical parameters for all extragalactic ($z>10^{-3}$) targets in DR1 observed by DESI (\citealt{moustakas23b}; Moustakas et~al. 2026, in preparation).\footnote{\url{https://fastspecfit.readthedocs.io}} It uses physically motivated stellar population synthesis and emission-line templates to  model each DESI spectrum jointly with the optical through infrared broadband photometry. The measurements included in this VAC include: stellar velocity dispersions; stellar masses; K-corrections and rest-frame colors and magnitudes; emission-line velocity widths, velocity shifts, fluxes, and equivalent widths (separately both narrow and broad emission-line components); and much more.

\underline{Extended Halo-based Group VAC} is the updated version of the extended halo-based group catalog derived from the Legacy Surveys DR9 for a galaxy sample limited to $z$-band apparent magnitude $z < 21$. For all the DESI target classes included in DR1, the photometric redshifts will be updated to spectroscopic redshifts. \citet{yang21a} provide more details regarding the extended halo-based group finder and sample selection.

\underline{Mass EMLines VAC} provides stellar mass and emission line measurements for all galaxies in DESI DR1 with reliable redshift measurements. Stellar masses are derived using CIGALE \citep{boquien19a}, which employs the broadband $g$-, $r$-, $z$-, $W1$-, and $W2$-band photometry from the Legacy Surveys, and spectrophotometry of 10 artificial bands generated through convolution with DESI spectra. A main set of optical emission lines are measured by a single Gaussian fit, with absorption correction through stellar continuum fitting performed using STARLIGHT\footnote{\url{http://www.starlight.ufsc.br}} \citep{cid-fernandes05a, cid-fernandes11a}. Additionally, the catalog includes stellar population properties derived by CIGALE and those obtained by STARLIGHT using DESI spectra. For additional information see \citet{zou24b} and Zou et~al. (2026, in preparation). 

\underline{Strong Lensing VAC} provides a comprehensive catalog of spectroscopic observation of strong gravitational lenses observed by DESI. The observations are primarily obtained through a program that seeks to obtain spectroscopic redshifts for both the lensing galaxies and the lensed sources for $\sim$1800 new strong lensing candidate systems (\citealt{huang25a}; Storfer et~al. 2025, in preparation). Objects included in the catalog were visually inspected in order to make a determination on the quality of the Redrock redshift, and to determine the nature of the system. Systems with at least one confirmed redshift for either the lens or source are included in the VAC. In addition to relevant data from the default DESI pipeline, the VAC includes data from the FastSpecFit VAC, including velocity dispersion, star-formation rate, and stellar mass. Also included is $1\farcs5$ diameter aperture photometry at the location of the fibers targeting lenses and sources. Lastly, with a number of Ly$\alpha$ emitters and absorbers not currently fit for redshifts via Redrock, we use custom template-fitting within the Redrock framework to fit for our own redshifts. Systems that are assigned redshifts this way are noted in the catalog. 

\subsection{Quasar Science VACs}\label{sec:vac_quasar}

\underline{AGN/Galaxy Classification VAC} provides AGN and QSO identification for galaxies ($z>0.001$) from all target classes observed within DR1 for objects with at least one DESI spectrum. Identification of QSOs and AGN is achieved through multiple means, including optical and ultraviolet emission-line diagnostics and multi-wavelength photometry. The VAC enables rapid selection of more complete versus more pure methodologies, as well as selection through any one AGN diagnostic (Juneau et~al. 2026, in preparation). A tutorial accompanies the catalog which demonstrates both how to use the catalog for rapid AGN / QSO selection as well as using the open source diagnostic code to select a specific line diagnostic and to choose the dividing line methodologies of those diagnostics. An example might be to test the [\ion{S}{2}] \citet{baldwin81a} classification assuming the more recent dividing lines of \citet[][implemented by default]{law21a} compared with the traditional dividing lines of \citet{kewley01b}, which were reported to potentially overestimate the presence of AGN \citep[e.g.,][]{pucha25a}.

\underline{AGN Host Properties VAC} provides physical properties for approximately 17M galaxies derived through spectral energy distribution fitting using CIGALE
\citep[v.22.1;][]{boquien19a}. This analysis accounts for contributions from both stars and, if present, an AGN. CIGALE uses the principle of energy balance, where dust-absorbed stellar emission in the ultraviolet and optical bands is re-emitted in the infrared; it estimates key galaxy and AGN properties such as stellar mass, star formation rate, and the AGN fraction (the contribution of the AGN dusty torus to the total IR luminosity \citep[e.g.,][]{ciesla15a, salim18a, yang20a, yang22a}. A detailed description of the catalog and its statistical properties based on data in the DESI EDR can be found in \cite{siudek24a}, with additional follow-up analysis using spectral energy distribution modeling to identify AGN to be described in \citet{siudek25a}.

\underline{BHMass VAC} provides \ion{Mg}{2}-based supermassive black hole mass estimates for 490,648 quasars at $0.6<z<1.6$ from DESI DR1. Approximately 35\% of the sample have mass uncertainties less than 0.5~dex. The iron-corrected mass is estimated using the fitting process and estimator introduced by \cite{pan25a}, which accounts for the effects of the Eddington ratio. Additionally, we integrate fitting results from the FastSpecFit VAC and offer alternative mass estimates based on the formulations by \citet{shen11a, le20a, yu23a}, providing users with flexibility in choosing their preferred estimator. This catalog will be periodically updated and extended to include more DESI quasars.

\underline{\ion{C}{4} Absorber VAC} contains the largest catalog to date of \ion{C}{4} absorption systems detected in quasar spectra from DESI DR1 \citep{anand25a}. The catalog includes $32,321$ \ion{C}{4} absorber systems identified along $94,986$ quasar sightlines over the redshift range $1.4 < z < 4.5$. Absorbers are detected using an automated matched-kernel convolution technique combined with adaptive signal-to-noise thresholds. Quasar continua are modeled using a non-negative matrix factorization (NMF) approach. Rest-frame equivalent widths are measured for both doublet components using a double-Gaussian profile fitting method, and column densities are derived using the apparent optical depth method \citep{savage91a}. Catalog completeness is quantified via Monte Carlo absorber-injection simulations, with the sample reaching approximately 50\% completeness at $\mathrm{EW} \gtrsim 0.4\,\AA$, while the overall catalog purity exceeds 95\%.

\underline{DLA NN and GP Finder VAC} contains the catalog of DLAs used for the DR1 Ly$\alpha$ BAO analysis. The catalog results from the combination of two DLA finding algorithms. The first one is based on a convolution neural network  trained with mock DESI spectra whose algorithm and performance is described in \citet{wang22a}. The second algorithm, which uses a Gaussian Process (GP), is based on the GP model of \cite{ho20a}. Both outputs from these algorithms are retained in the combined catalog and DLA candidates are merged when they are found by both methods with a velocity difference of less than 800~km~s$^{-1}$. The merged catalog includes estimated redshifts and \ion{H}{1} column densities together with confidence flags from both algorithms. It contains 54,416 candidate DLAs with \ion{H}{1} column density $>10^{20.3}$~cm$^{-2}$ within Ly$\alpha$ forest spectra with signal-to-noise ratio larger than three. 

\underline{DLA Template Finder VAC} contains candidate DLAs identified with the DLA Toolkit software, which uses spectral template-fitting. The catalog was constructed from quasar spectra with $2.0<z<4.25$, allowing for a maximum of three detections per sight-line; it contains an estimated DLA redshift, \texttt{HI} column density, and detection significance for each DLA. \citet{brodzeller25a} provide the full details regarding the construction of the catalog, parameter accuracy, and recommended quality cuts to maximize sample purity and completeness. This catalog is supplemental to the DLA NN and GP Finder VAC.

\underline{\ion{Mg}{2} Absorber VAC} contains information regarding the detection and characterization of \ion{Mg}{2} absorption systems in DESI quasar spectra. The catalog is based on a search for \ion{Mg}{2} absorbers in spectra from all quasar targets.
Our search technique employs both a non-negative matrix factorization continuum construction step, which identifies possible doublet candidates, as well as an Markov Chain Monte Carlo-based line-fitting step which determines accurate line-statistics. From a parent sample of 1.47M DESI DR1 quasars we detect a total of 270,529 \ion{Mg}{2} absorbers with velocity offsets from the background quasar of $>-5000$ km~s$^{-1}$, and 392 \ion{Mg}{2} absorbers with velocity offsets $<-5000$ km~s$^{-1}$. 
From an analysis of results from DESI EDR, we estimate this catalog to have a purity greater than 99\% and a completeness greater than 90\% for absorbers with rest-frame equivalent widths greater than 0.8~\AA.

\underline{ZLyA VAC} contains the redshifts and broad absorption line (BAL) information used for the DR1 Ly$\alpha$ forest BAO analysis \citep{DESI2024.IV.KP6} The catalog was constructed with $z>1.6$ main-survey, dark-program quasars with a Redrock \texttt{ZWARN} flag of either zero or four (see \S\ref{sec:redrock}).\footnote{\url{https://desidatamodel.readthedocs.io/en/latest/bitmasks.html\#zwarn}}, Redshifts were updated from the LSS catalog redshifts to correct for a systematic underestimation of quasar redshifts introduced by improper handling of Lyman-series optical depth \citep{RedrockQSO.Brodzeller.2023, KP6s4-Bault}. The new redshifts were determined using a modified version of the Redrock \texttt{HIZ} quasar templates that incorporate the \citet{kamble20a} model for Ly$\alpha$ effective optical depth redshift evolution. BAL attributes were derived using the method presented in \citet{filbert24a}.

\subsection{Lyman-Alpha Forest VACs}\label{sec:vac_lya}

\underline{Lyman Alpha Forest Year 1 Deltas VAC} contains the flux-transmission field used in the DR1 Ly$\alpha$ BAO measurement. The fluctuations in the calibration region (rest-frame wavelength from 1600 to 1850 \AA), the Ly$\alpha$ region A (rest-frame wavelength from 1040 to 1205 \AA), and Ly$\beta$ region B (rest-frame wavelength from 920 to 1020 \AA) were obtained using Python Code for IGM Cosmological-Correlations Analyses (PICCA; \citealt{du-mas-des-bourboux20a}). The continuum-fitting process is discussed in detail by \citet{ramirez-perez24a}. This catalog provides the estimated quasar continua, the fluctuations in the spectra, and their associated weights for all lines-of-sight for each HEALPix pixel. The fitted variance functions, stack of fluctuations, and the mean continuum are stored for each iteration during the continuum-fitting process.

\underline{Lyman Alpha Forest Year 1 Correlations VAC} contains the correlation data products that are used in the DR1 Ly$\alpha$ BAO measurement. It contains correlation measurements between Ly$\alpha$ forest fluctuations in both regions A (1040 to 1205 \AA) and B (920 to 1020 \AA), and tracer quasars. For each of the four combinations, we provide the correlation function and the distortion matrix. The latter is used to forward-model the effect of reconstructing the unabsorbed flux of quasar spectra (see, e.g., \citealt{du-mas-des-bourboux20a}). Furthermore, we provide one smoothed covariance matrix that includes the cross-covariances between the different correlation function measurements.

\bibliographystyle{aasjournal}
\bibliography{DESI_supporting_papers, dr1}

\end{document}